\documentclass[12pt,twoside]{article}
\usepackage{amssymb}
\usepackage{amsmath}
\usepackage{latexsym}
\usepackage{longtable}
\usepackage{epsfig}
\usepackage{graphicx,bbm,psfrag}
\graphicspath{{images/}}

\begin{document}
\begin{titlepage}
\begin{center}
\vspace*{2cm} {\Large {\bf Nonlinear Fluctuating Hydrodynamics for Anharmonic
Chains\bigskip\bigskip\\}}
{\large Herbert Spohn}\bigskip\bigskip\\
Institute for Advanced Study, Princeton, NJ 08540,\\ and\\
  Zentrum Mathematik, Physik Department, TU M\"unchen,\\
 Boltzmannstr. 3, D-85747 Garching, Germany\\e-mail:~{\tt spohn@tum.de}\\

\end{center}
\vspace{5cm} \textbf{Abstract.} With focus on anharmonic chains, we develop a nonlinear version of fluctuating hydrodynamics, in which the Euler currents are kept to second order in the deviations from equilibrium and dissipation plus noise are added. The required model-dependent parameters are written in such a way that they can be computed numerically within seconds, once the interaction potential, pressure, and temperature are given. In principle the theory is applicable to any one-dimensional system with local conservation laws. The resulting nonlinear stochastic field theory is handled in the one-loop approximation. Some of the large scale predictions can still be worked out analytically. For more details one has to rely on numerical simulations of the corresponding mode-coupling equations. In this way we arrive at detailed predictions for the equilibrium time correlations of the locally conserved fields of an anharmonic chain.
\end{titlepage}

\section{Introduction}\label{sec1}
\setcounter{equation}{0}

Fermi, Pasta, and Ulam investigated the approach to thermal equilibrium choosing as test case a chain of anharmonicly coupled particles. While this enterprise turned out to be a more complicated issue than anticipated, lasting results are spin-offs like solitons, breathers, and the approximation of the FPU dynamics through integrable systems. We refer to~\cite{FPU1,
FPU2} for a modern perspective. The non-equilibrium statistical mechanics of anharmonic chains was lingering until in late 90ies when in numerical simulations clear evidence of anomalous energy transport was discovered~\cite{Livi97}. While some aspects of this phenomenon were noted earlier, only then systematic investigations started to be carried out by many groups, see the most informative reviews~\cite{LeLi03,Dh08}. Even today the situation is not settled and further molecular dynamics simulations are in progress.

The main goal of my contribution is to explain nonlinear fluctuating hydrodynamics when applied to the equilibrium time correlations  of the conserved fields for anharmonic chains. We will also discuss the analytical predictions based on a mode-coupling approximation. To my own surprise, the basic theory of anharmonic chains is not well covered in the literature. For example, experts in the field do not seem to be aware of the fact that there is a simple, explicit formula for the sound speed, valid in complete generality. Hence, we will have start from scratch, emphasizing that structurally the chains  can be viewed
as a 1+1 dimensional field theory.

The research presented is  greatly inspired by the recent letter~\cite{vB12} of Henk van Beijeren. Our main novel point is to work out nonlinear fluctuating hydrodynamics as a mesoscopic self-contained description. This theory
can be applied also to other systems, for example one-dimensional quantum fluids, quantum spin chains, and multi-component lattice gases in one dimension. Also, on the level of nonlinear fluctuating hydrodynamics one can easily include boundary reservoirs to drive the system towards a non-equilibrium steady state, which is of independent interest.

Our contribution has a slightly unusual format. The background for anharmonic chains is explained in Sec.~\ref{sec2} and  nonlinear fluctuating hydrodynamics is developed systematically in Sec.~\ref{sec3}. Nonlinear fluctuating hydrodynamics is a stochastic field theory which will be handled in the one-loop approximation. The analytic predictions based on the corresponding mode-coupling equations are presented in Secs.~\ref{sec4} - \ref{sec6}.   A detailed account on previous related work, including the comparison with available data from molecular dynamics, is delayed to Sec.~\ref{sec7}. There one finds also a discussion of stochastic models. While they are outside of our main frame, these models provide valuable material on the validity of the stochastic field theory.
To make the text more readable, many computations and details are shifted to the Appendices. In Appendix \ref{app.A} we provide the nonlinear coupling coefficients as expressed in terms one-dimensional integrals. These coupling parameters are
the quantitative bridge from the microscopic to the mesoscopic description. In Appendix \ref{A.h} the hard-point gas with alternating masses is discussed, both as an illustration of the method and as a numerically well studied system. 
The derivation of the mode-coupling equations is presented in Appendix \ref{app.B}.

The mode-coupling equations are approximate but still sufficiently simple to be solved by a numerical iteration
technique. Based  on coupling coefficients from particular anharmonic chains, Christian Mendl has performed many simulations,  which are highly instructive for a better understanding of the intermediate time regime and the dependence on model parameters.
A short summary is provided in~\cite{MS}. In this contribution we do not touch upon the numerics.

\section{Dynamics, equilibrium states, Euler equations}\label{sec2}
\setcounter{equation}{0}

We consider $N$ particles in one dimension with positions, $q_j$, momenta $p_j$, $j=1,\ldots,N$. Their mass is set equal to one. Nearest neighbors in index space are coupled through a potential $V(y)$, which is bounded from below and increases at infinity, at least linearly when either $y\to\infty$ or $y\to-\infty$. The hamiltonian of the chain thus reads
\begin{equation}\label{1}
  H=\sum^{N}_{j=1} \big(\tfrac{1}{2}p^2_j+V(q_{j+1}-q_j)\big)\,,
\end{equation}
implying the equations of motion
\begin{equation}\label{1a}
\frac{d^2}{dt^2}q_j(t)  =  V'(q_{j+1}(t) - q_{j}(t)) - V'(q_{j}(t) - q_{j-1}(t))\,.
\end{equation}

We enforce the periodic boundary condition
\begin{equation}\label{2}
  q_{N+1}=q_1+L
\end{equation}
for a chain of length $L$. Note that there is no strict spatial ordering of particles, in general. Also $L$ can have either sign. For $L\ll -1$, the $N$-th particle will tend to be at the left-most location. Instead of the microcanonical constraint~(\ref{2}), another natural boundary condition would be to pull/push particle 1 and $N$ with a linear force. This modification will not be considered and we stick to~(\ref{2}).

Eqs.~(\ref{1a}) and~(\ref{2}) suggest two distinct physical interpretations. Firstly one could start from the nonlinear wave equation
\begin{equation}\label{2a}
\partial^2_tu(x,t) = \partial_xV'(\partial_x u(x,t))
\end{equation}
with $u(x,t)$ the field amplitude at location $x$ and time $t$. In a  lattice discretization with lattice spacing 
$\epsilon$, where $q_j(t) = u(x,t)$ with $\epsilon j = \lfloor x\rfloor_\epsilon$,  $\lfloor \cdot \rfloor_\epsilon$ denoting integer part mod $\epsilon$,
one arrives at~(\ref{1a}). The boundary condition~(\ref{2}) imposes a slope $L/N$
in $u$. Thus~(\ref{1a}) is regarded as a one-dimensional lattice field theory, which is the point of view adapted throughout this text.
In the second interpretation one regards the system as a fluid. Then physical space is the real line, labeled particles move on the line and interact with their nearest neighbor particles in index space 
through some nonlinear spring. The initial condition is periodic in the sense that $q_j = q_{j+N}$, a property which is preserved in the course of time. One could also consider particles on a circle of length $L$. Then the springs could in principle wrap around the circle and one has to keep track of the  particles' winding numbers. Such unphysical features disappear in the infinite volume limit, which will anyhow be considered later on.

It will be convenient to introduce the positional differences 
\begin{equation}\label{3}
r_j=q_{j+1}-q_j\,,
\end{equation}
more physically referred to also as compressions, resp. elongations. If ordered, $r_j$ is  also the free volume between 
particle $j+1$ and its left neighbor $j$. But such terminology would be less descriptive and we stick to elongation field. The equations of motion (\ref{1a}) then become
\begin{equation}\label{4}
\dot{r}_j=p_{j+1}-p_j\,,\quad p_{N+1}=p_1\,,
\end{equation}
\begin{equation}\label{5}
\dot{p}_j=V'(r_j)-V'(r_{j-1})\,,\quad r_0=r_N\,,\quad j=1,\ldots,N\,.
\end{equation}
(\ref{4}) is viewed as a classical  lattice field theory or ``spin'' system. The underlying lattice is $[1,\ldots,N]$ with periodic boundary conditions. The field variables are $(r_j,p_j)_{j=1,\ldots,N}$. Through~(\ref{4}) there is a coupling to the right neighbor and through~(\ref{5}) a coupling to the left neighbor.

Since
\begin{equation}\label{6}
\frac{d}{dt}\sum^N_{j=1} r_j=0\,,\quad \frac{d}{dt}\sum^N_{j=1} p_j=0\,,
\end{equation}
one may impose the constraints of total length and total momentum,
\begin{equation}\label{7}
\sum^N_{j=1} r_j=L\,,\quad \sum^N_{j=1} p_j=0\,.
\end{equation}
The first constraint is a transcription of~(\ref{2}), while zero total momentum can always be achieved by a uniform shift of all $p_j$. Also the dynamics does not change under a simultaneous shift of $V(y)$ to $V(y-a)$ and $r_j$ to $r_j+a$.
In this sense the potential is defined only up to translations.

For the microcanonical equilibrium state one uses extensive constraints for elongation, momentum, and energy as
\begin{equation}\label{8}
N\ell=\sum^N_{j=1} r_j\,,\quad \sum^N_{j=1} p_j=0\,,\quad N \mathsf{e}=\sum^N_{j=1} \big(\tfrac{1}{2}p^2_j+ V(r_j)\big)
\end{equation}
with $\ell$ the elongation per particle and $\mathsf{e}$ the internal energy per particle. The microcanonical state is then the uniform measure under these constraints. For the energy, it will be convenient to smoothen the sharp constraint~(\ref{8}) to a small energy interval. In our context equivalence of ensembles holds and computationally it is of advantage to switch to the canonical ensemble with respect to all three constraints. In the fluid literature this is also referred to as pressure ensemble. Then the dual variable for the elongation $\ell$ is the pressure $p$ and for the internal energy $\mathsf{e}$ the inverse temperature $\beta$. Under the canonical ensemble the collection $(r_j,p_j)_{j\in\mathbb{Z}}$ are independent random variables. Their single site probability density is given by
\begin{equation}\label{9}
Z^{-1} \mathrm{e}^{-\beta(V(y)+py)} (2\pi/\beta)^{-1/2} \mathrm{e}^{-\beta v^2/2}\,,\quad Z(p,\beta)=\int_\mathbb{R}dy \mathrm{e}^{-\beta(V(y)+py)}\,.
\end{equation}
Averages with respect to~(\ref{9}) are denoted by $\langle\cdot\rangle_{p,\beta}$. Note that
\begin{equation}\label{10}
  p=-\langle V'(y)\rangle_{p,\beta}
\end{equation}
and, as average force on a specified particle, $p$ is identified with the thermodynamic pressure. From~(\ref{9}) one deduces that $\beta > 0$, while $p\in \mathbb{R}$ in case $V$ increases faster than linearly at $\pm \infty$. For a slower increase, the range of $p$ has to be restricted. One famous example is the Toda chain with $V(y) = \mathrm{e}^{-y}$, in which case $p>0$ is required in order to have a finite partition function. The canonical free energy is given by
\begin{equation}\label{11}
G(p,\beta)=-\beta^{-1} \big(-\tfrac{1}{2}\log\beta + \log Z(p,\beta)\big)\,.
\end{equation}
Then
\begin{equation}\label{12}
\ell =\langle y\rangle_{p,\beta}\,,\quad \mathsf{e}=\partial_\beta\big(\beta G(p,\beta)\big) -p \ell=\frac{1}{2\beta}+\langle V\rangle_{p,\beta}\,.
\end{equation}
(\ref{12}) defines $(p,\beta) \mapsto (\ell(p,\beta),\mathsf{e}(p,\beta))$, thereby the inverse map 
$(\ell, \mathsf{e}) \mapsto (p(\ell,\mathsf{e}),$ $  \beta(\ell,\mathsf{e}))$, and thus accomplishes the switch between
 the microcanonical variables $\ell, \mathsf{e}$ and the canonical variables $p, \beta$.
The canonical state has zero average momentum. If instead of~(\ref{7}) one imposes an extensive non-zero momentum constraint, 
\begin{equation}\label{7a}
 \sum^N_{j=1} p_j=N\mathsf{u}\,,
\end{equation}
then the Maxwellian is shifted by $\mathsf{u}$ and
$\mathsf{e}$ turns into  the total energy per particle, denoted by $\mathfrak{e}$, with $\mathfrak{e} = \tfrac{1}{2}\mathsf{u}^2 +\mathsf{e}$.

The next basic information on the model are the local conservation laws. From~(\ref{5}) one deduces immediately that elongation and momentum are locally conserved.  The corresponding local currents are $-p_j$ and $-V'(r_{j-1})$. As for any mechanical system also the local energy is conserved. The energy at site $j$ is
\begin{equation}\label{13}
e_j=\tfrac{1}{2}p^2_j + V(r_j)\,.
\end{equation}
Then
\begin{equation}\label{14}
\dot{e}_j= p_{j+1} V'(r_j)-p_j V'(r_{j-1})\,,
\end{equation}
hence the local energy current is $-p_j V'(r_{j-1})$. We collect the conserved fields as the $3$-vector
$\vec{g} = (g_1,g_2,g_3)$, 
\begin{equation}\label{14c}
\vec{g}(j,t) = \big(r_j(t),p_j(t),e_j(t)\big) \,,
\end{equation}
$\vec{g}(j,0) = \vec{g}(j)$. Then
\begin{equation}\label{14a}
\frac{d}{dt}\vec{g}(j,t) + \vec{\mathcal{J}}(j+1,t)  - \vec{\mathcal{J}}(j,t)=0 \,,
\end{equation}
where the local current functions are given by
\begin{equation}\label{14b}
\vec{\mathcal{J}}(j) = \big( -p_j,-V'(r_{j-1}), - p_jV'(r_{j-1})\big)\,.
\end{equation}
We note that there is some arbitrariness of how to split the potential energy between neighboring particles. For fluids the conventional choice is to split the potential energy in half. In our context this would lead to conserved fields which in equilibrium are
correlated in index space, which is not so convenient.

Once the conserved fields are identified, the next step is to consider initial states where the parameters $\ell, \mathsf{u},
\mathfrak{e}$, equivalently $p,\beta$ and mean velocity, are slowly varying on the scale of the lattice. This is
 the hydrodynamic approximation. If we use $x\in\mathbb{R}$ for the corresponding continuum approximation the conserved fields are the local elongation $\ell(x)$, the local momentum $\mathsf{u}(x)$, and the local total energy $\mathfrak{e}(x) =\frac{1}{2} \mathsf{u}(x)^2+ \mathsf{e}(x)$ per particle. The fields evolve slowly in time and preserve local equilibrium, at least approximately. Hence, by averaging the fields in a local equilibrium state,  the microscopic conservation laws turn into the Euler equations of an anharmonic chain as
\begin{equation}\label{15}
\partial_t\ell +\partial_x \mathsf{j}_\ell =0\,,\quad   \partial_t \mathsf{u} +\partial_x \mathsf{j}_\mathsf{u} =0\,,\quad   \partial_t \mathfrak{e} +\partial_x \mathsf{j}_\mathfrak{e} =0\,,
\end{equation}
where the hydrodynamic currents are given by
\begin{equation}\label{16}
\langle \vec{\mathcal{J}}(j)\rangle_{\ell,\mathsf{u},\mathfrak{e}} = \big(-\mathsf{u},p(\ell,\mathfrak{e}-\tfrac{1}{2}\mathsf{u}^2), \mathsf{u} p(\ell,\mathfrak{e}-\tfrac{1}{2}\mathsf{u}^2)\big)
= \vec{\mathsf{j}}
\end{equation}
with $p(\ell,\mathsf{e})$ defined implicitly through~(\ref{12}). 

\section{Dynamic correlations of the conserved fields, nonlinear fluctuating hydrodynamics}\label{sec3}
\setcounter{equation}{0}

The three hydrodynamics fields are the slowest degrees of freedom of the chain and their equilibrium time correlations are of central interest. There are other slow modes, like products of conserved fields or currents, but for us here of secondary interest. Taking already the limit of an infinitely long chain, the covariance  matrix of the conserved fields is defined by
\begin{equation}\label{17}
S_{\alpha\alpha'}(j,t)=\langle g_{\alpha}(j,t) g_{\alpha'}(0,0)\rangle_{p,\beta} - \langle g_{\alpha}(0)\rangle_{p,\beta} \langle g_{\alpha'}(0)\rangle_{p,\beta}\,,
\end{equation}
$\alpha,\alpha'=1,2,3$. By space-time stationarity
\begin{equation}\label{20b}
S_{\alpha\alpha'}(j,t)= S_{\alpha'\alpha}(-j,-t)\,.
\end{equation}
Hence in the following we will assume $t \geq 0$ throughout. We also use the notation $S(j,t)$ for the $3\times 3$ matrix in~(\ref{17}). $\langle\cdot\rangle_{p,\beta}$ is the equilibrium state, which by assumption has zero average momentum and is simply a product measure with single site density~(\ref{9}). Of course, one could consider also the finite $N$ version, in which case one would have to distinguish between microcanonical and canonical equilibrium state. In principle, the microcanonical state is to be preferred, since then the limit $t\to\infty$ in~(\ref{17}) will vanish because of time mixing. On the other hand the canonical state is easily generated as a family of independent random variables and one does not have to rely on equilibration through the dynamics, which could be a source of error. At $t=0$ the field components are uncorrelated in $j$, \textit{i.e.}
\begin{equation}\label{19}
S(j,0)=\delta_{j0} C
\end{equation}
with the static susceptibility matrix
\begin{equation}\label{20}
C=
\begin{pmatrix} \langle y;y\rangle_{p,\beta} & 0 & \langle y;V\rangle_{p,\beta} \\
                0 & \beta^{-1} & 0 \\
                \langle y;V\rangle_{p,\beta} & 0 & \frac{1}{2}\beta^{-2}+\langle V;V\rangle_{p,\beta}
\end{pmatrix}\,.
\end{equation}
Here $\langle X;Y\rangle=\langle XY\rangle-\langle X\rangle\langle Y\rangle$ denotes the second cumulant.
The conservation law implies the sum rule
\begin{equation}\label{20a}
\sum_{j \in \mathbb{Z}}S_{\alpha\alpha'}(j,t) = C_{\alpha\alpha'}\,.
\end{equation}

As a general insight, one has to treat all conserved fields on the same footing. The time-dependent
statistical correlations along the chain can be captured by understanding the interaction between the conserved modes. Just to treat one single conserved field is likely to miss some of these dynamical processes.

(\ref{17}) may be viewed as the average fields at time $t$, given that there is initially a small perturbation at $j=0$. Hence, in roughest approximation, the correlator $S$ should be governed by the Euler equations linearized as $\ell+u_1(x)$, $0+u_2(x)$, $\mathsf{e}+u_3(x)$ to linear order in $\vec{u}(x)$, in other words by the linear equation
\begin{equation}\label{21}
\partial_t \vec{u}(x,t)+\partial_x A \vec{u} (x,t)=0
\end{equation}
with
\begin{equation}\label{22}
A=
\begin{pmatrix} 0 & -1 & 0 \\
               \partial_\ell p & 0 & \partial_\mathsf{e}p \\
                0 & p & 0
\end{pmatrix}\,.
\end{equation}
Following the standard thermodynamic conventions, $p=p(\ell,\mathsf{e})$  in~(\ref{22}). For the numerical implementation of our theory, it will be convenient to reexpress $A$ in canonical variables. Not to interrupt the main argument, the relevant formulas will be summarized in the Appendix \ref{A.d}. 
Beyond~(\ref{20a}) there is the second sum rule
\begin{equation}\label{22a}
\sum_{j \in \mathbb{Z}}jS(j,t) = AC\,t\,,
\end{equation}
see Appendix \ref{A.f}. Hence, either using~(\ref{20b}) or by explicit computation,
\begin{equation}\label{23}
  AC=CA^{\mathrm{T}}\,,
\end{equation}
where $^\mathrm{T}$ denotes transpose. Since $C>0$, $A$ has real eigenvalues which turn out to be equal to  
$0,\pm c$ with $c$ the sound speed given by
\begin{equation}\label{25}
c^2= -\partial_{\ell} p+p \partial_\mathsf{e}p\,.
\end{equation}

In what follows it will be important to make a linear transformation such that the components in Eq. (\ref{21})
decouple and each one evolves with a definite velocity. The corresponding linear transformation will be denoted by $R$,
which acts in component space only. Hence $\vec{\phi} = R\vec{u}$ and $R$ has to satisfy
\begin{equation}\label{24}
RAR^{-1}= \mathrm{diag}(-c,0,c)\,.
\end{equation}
By convention, $ \vec{\phi} $ is referred to as normal modes. $\vec{u}(x,t)$ are the physical fields and
$\vec{\phi}(x,t)$ are the normal mode fields. One can also transform the lattice field $\vec{g}(j,t)$ to $R\vec{g}(j,t)$.
Then the correlator is transformed as
\begin{equation}\label{26a}
S_{\alpha\alpha'}^\sharp(j,t) = (RSR^{\mathrm{T}})_{\alpha\alpha'}(j,t) = \langle (R\vec{g})_{\alpha}(j,t) ;(R\vec{g})_{\alpha'}(0,0)\rangle_{p,\beta} \,.
\end{equation}
Here the superscript $^\sharp$ is used throughout to signal normal mode covariances. Secondly, one requires
$R\vec{g}(j)$  to be uncorrelated in equilibrium, both in lattice site and index, which means
\begin{equation}\label{24a}
RCR\mathrm{^T}=1\,.
\end{equation}  
As discussed in Appendix \ref{A.b}, the matrix $A$ has a system of left and right eigenvectors through which  the $R$-matrix can be computed. Up to a global sign the linear transformation $R$ is unique.

At the current linearized level of approximation the correlator $S^\sharp$ is diagonal and satisfies
\begin{equation}\label{26}
\partial_t S^\sharp_{\alpha\alpha}(j,t)+c_\alpha(S^\sharp_{\alpha\alpha}(j+1,t)- S^\sharp_{\alpha\alpha}(j,t))=0
\end{equation}
with $S^\sharp_{\alpha\alpha}(j,0)= \delta_{0j}$. Thus $S_{\alpha\alpha'}(j,t)$ has three sharp peaks
as a function of $j$, the two sound mode peaks located at $j=\pm ct$ and the heat mode peak located at $j=0$. For the exact $S(j,t)$ the peaks will be broadened because of dissipation plus noise, which can be captured only by a more elaborate theory.

For fluids in physical space, the standard procedure, see~\cite{Forster,LL,RdeL}, is to lift the Euler equations to the compressible Navier-Stokes equations, which include friction by second order derivative terms. Their strengths are 
expressed through the transport coefficients, namely thermal conductivity and bulk viscosity. On the linearized level this provides the desired broadening. As one basic principle of statistical mechanics, according to Onsager, dissipation is always connected with fluctuations. Thus the appropriate dynamical evolution equations, at this level of precision, are the linearized Navier-Stokes equations with added  random currents, which are modeled as space-time white noise of a strength fixed by the fluctuating-dissipation relation. This theory is known as linear fluctuating hydrodynamics.

On a sufficiently coarse space-time scale the theory is accurate and covers also non-equilibrium steady states, see~\cite{Tr} for a more complete discussion. For anharmonic chains, and other momentum conserving one-dimensional systems, linear fluctuating hydrodynamics unfortunately fails, since as a rule the total current-current correlations decay non-integrably,
which makes  the 
Green-Kubo formula for the transport coefficients to diverge. On the level of equilibrium time correlations, this means that the sound mode and heat mode peaks broaden super-diffusively. For a quantitative description we postulate that a minimal nonlinear extension of fluctuating hydrodynamics suffices.

Our starting point is still standard linear fluctuating hydrodynamics, which in our context is the Langevin equation
\begin{equation}\label{28}
\partial_t \vec{u}(x,t) +\partial_x \big(A \vec{u}(x,t) -\partial_x \tilde{D} \vec{u}(x,t) +\tilde{B}\vec{\xi}(x,t)\,\big)=0\,.
\end{equation}
Here $\tilde{D}$ is the diffusion matrix, $\tilde{D}=\tilde{D}\mathrm{^T}$, $\tilde{D}>0$, $\vec{\xi}$ Gaussian white noise with mean 0 and covariance
\begin{equation}\label{29}
\langle\xi_\alpha(x,t) \xi_{\alpha'}(x',t')\rangle= \delta_{\alpha\alpha'} \delta(x-x') \delta(t-t')\,,
\end{equation}
and $\tilde{B}\tilde{B}\mathrm{^T}$ the noise strength matrix.  $\vec{u} (x,t)$ is now elevated to a Gaussian process. Since 
$\vec{u} (x,t)$ models the deviations from uniformity, we will consider the mean zero stationary process $\vec{u} (x,t)$ governed by~(\ref{28}). At fixed time $t$ its spatial statistics is mean zero white noise with covariance
\begin{equation}\label{30}
\langle u_\alpha(x,t) u_{\alpha'}(x',t)\rangle =\tilde{C}_{\alpha\alpha'} \delta (x-x')\,,
\end{equation}
where the susceptibility matrix $C$ and $\tilde{D}$, $\tilde{B}$ satisfy the fluctuation-dissipation relation
\begin{equation}\label{31}
  \tilde{D}\tilde{C}+\tilde{C}\tilde{D}=\tilde{B}\tilde{B}^\mathrm{T}\,.
\end{equation}
In equilibrium the conserved fields are uncorrelated in $j$, which viewed on a coarse scale turns into a Gaussian
fluctuation field with covariance $C_{\alpha\alpha'}\delta(x-x')$.  Hence $\tilde{C}$ in~(\ref{30}) is chosen to be identical to the covariance matrix $C$ of~(\ref{19}). On the other hand, because of super-diffusive broadening, there seems to be no basic principle how to determine $\tilde{D}$ and $\tilde{B}$ separately.~(\ref{28}) holds only on a somewhat vaguely defined mesoscopic scale.

Nonlinear fluctuating hydrodynamics claims that in order to capture the super-diffusive broadening it suffices to expand the Euler equation up to second order. It is still under debate whether such a minimal modification of~(\ref{28}) is a valid approximation of the true $S(j,t)$. In fact, molecular dynamics simulators are not so convinced. On the other hand, so far, nonlinear fluctuating hydrodynamics has presented only asymptotic results for which a serious comparison has turned out to be difficult. One goal of our undertaking is to provide the full space-time dependence of the equilibrium time correlations of the conserved fields as derived from nonlinear fluctuating hydrodynamics.
This information should facilitate the comparison with molecular dynamic simulations.

As to be  stressed, the validity of~(\ref{28})  and its nonlinear version requires the absence of additional conservation laws and the dynamics to be sufficiently chaotic so as to have good mixing in time.
These assumptions rule out the harmonic chain, $V(y) = \tfrac{1}{2}\omega_0^2y^2$, in an obvious way since the 
pressure is linear, $p = -\omega_0^2\ell$. The equal mass hard-point gas and the Toda chain fail more subtly. Both systems have an infinite number of conservation laws. But their macroscopic currents are nonlinear, hence nonlinear fluctuating hydrodynamics is non-degenerate. Still its predictions are not expected to match with the correlations of the mechanical system.

Expanding the currents to second order in $\vec{u}$ adds in~(\ref{28}) the term
\begin{equation}\label{32}
\tfrac{1}{2}\langle \vec{u}, \vec{H} \vec{u}\rangle =
\tfrac{1}{2}\sum^3_{\alpha,\beta=1}\vec{H}_{\alpha\beta} u_\alpha u_\beta
\end{equation}
in the bracket after $\partial_x$, where $H^\gamma$ is the Hessian matrix of $\mathsf{j}_\gamma$,
\begin{equation}\label{33}
  H^\gamma_{\alpha\beta} =\partial_{u_\alpha} \partial_{u_\beta} \mathsf{j}_\gamma
\end{equation}
evaluated at the equilibrium parameters $(\ell,0,\mathsf{e})$.

(\ref{28}) including the additional term~(\ref{32}) is transformed to normal modes $\vec{\phi}$ through $\vec{\phi} =R\vec{u}$ with the result
\begin{equation}\label{34}
\partial_t \phi_\alpha + \partial_x \big(c_\alpha \phi_\alpha + \langle\vec{\phi}, G^{\alpha}\vec{\phi}\rangle-\partial_x(D\phi)_\alpha+(B\xi)_\alpha\big)=0\,,
\end{equation}
where $ D = R \tilde{D}R^{-1}$ and $B = R\tilde{B}$ with noise strength $ B B^{\mathrm{T}} = 2 D$. $\alpha=-1,0,1$ and the velocity of the $\alpha$-th normal mode is $c_\alpha$, $c_\sigma=\sigma c$, $c_0=0$, $\sigma=\pm1$.  The inner product $\langle\cdot,\cdot\rangle$ is in component space and the $G^\alpha$ matrix of coefficients stands for
\begin{equation}\label{35}
 G^\alpha = \tfrac{1}{2}\sum^3_{\alpha'=1}  R_{\alpha\alpha'}  (R^{-1})^{\mathrm{T}} H^{\alpha'}R^{-1}\,.
\end{equation}
To achieve an understanding of the dynamics of the normal mode correlations, the main building blocks are the numerical values of the sound speed $c$ and the coupling matrices $G^\alpha$. Their computation is discussed in detail in Appendix \ref{A.b}, \ref{A.c}. Upon request, a Mathematica program is available which yields $c,G^\alpha$ for given $V,p,\beta$.

In equilibrium the fluctuations are stationary in space-time. Hence we consider the stationary process $\vec{\phi}(x,t)$ with mean zero, $\langle\vec{\phi}(x,t)\rangle = 0$,
defined through the solution of Eq.~(\ref{34}). The $\vec{\phi}$\hspace{1pt}-$\vec{\phi}$ correlator reads
\begin{equation}\label{35b}
S^{\sharp\phi}_{\alpha\alpha'}(x,t) = \langle \phi_{\alpha}(x,t)  \phi_{\alpha'}(0,0) \rangle\,,
\end{equation}
where the superscript $^\sharp$ reminds of normal mode and $^\phi$ of the underlying stochastic process. As our central conjecture
\begin{equation}\label{35e}
S^{\sharp}_{\alpha\alpha'}(j,t) \simeq S^{\sharp\phi}_{\alpha\alpha'}(x,t) \,,
\end{equation}
on a mesoscopic scale. As before, we specify a lattice spacing $\epsilon$ such that $\epsilon j = \lfloor x\rfloor_\epsilon$. More precisely, taking the appropriate scaling limit on \textit{both} sides of Eq.~(\ref{35e}),
one expects to achieve equality. Returning to the correlations of the physical fields, $S_{\alpha\alpha}(j,t)$, generically one would expect to have three peaks, the heat peak and two mirror  symmetric sound peaks. This is indeed valid with one exception. Since $R_{00} = 0 = (R^{-1})_{00}$, for $S_{22}(j,t)$ the heat peak vanishes always. 

To arrive at predictions for the anharmonic chains, this leaves us with the task to analyse the 
stationary two-point function of the stochastic field theory~(\ref{34}). But before, let us discuss the case of a single mode,
in which case Eq.~(\ref{34}) is the stochastic Burgers equation. In the frame moving with velocity $c_1$ it reads
\begin{equation}\label{36}
\partial_t\phi_1+\partial_x\big(G^1_{11}\phi^2_1 -\partial_x D\phi_1 +\sqrt{2D}\xi_1\big)=0\,.
\end{equation}
Introducing a height function, $h$,  as $\partial_x h=\phi_1$, Eq.~(\ref{36}) turns into the one-dimensional KPZ equation
\begin{equation}\label{37}
\partial_t h= - G^1_{11}(\partial_x h)^2 +D\partial^2_x h -\sqrt{2D}\xi_1\,,
\end{equation}
which has been studied in great detail over the recent years, see Section \ref{sec7}. In particular it has been proved that for the stationary $\phi_1(x,t)$ process, $\langle \phi_1\rangle=0$, the two-point function behaves for large $x,t$ as
\begin{equation}\label{38}
\langle\phi_1(x,t)\phi_1(0,0)\rangle\cong (\lambda_\mathrm{B} t)^{-2/3} f_{\mathrm{KPZ}}
\big((\lambda_\mathrm{B} t)^{-2/3}x\big)\,,
\end{equation}
where $\lambda_{\mathrm{B}}= 2\sqrt{2}|G^1_{11}|$. The universal scaling function $f_{\mathrm{KPZ}}$ is tabulated in~\cite{Prhp}, denoted there by $f$. $f_{\mathrm{KPZ}}\geq 0$, $\int dxf_{\mathrm{KPZ}}(x)=1$, $f_{\mathrm{KPZ}}(x)=f_{\mathrm{KPZ}}(-x)$, $\int dxf_{\mathrm{KPZ}}(x)x^2=0.510523\ldots$ . $f_{\mathrm{KPZ}}$ looks like a Gaussian with a large $|x|$ decay as $\exp[-0.295|x|^{3}]$~\cite{PrSp04}.


\section{Mode-coupling theory}\label{sec4}
\setcounter{equation}{0}

The nonlinear stochastic conservation laws~(\ref{34}) are complicated. For $n=1$ the one-loop approximation was written down in~\cite{vBKS85} with a numerical integration provided in~\cite{CM01}. The difference to the exact asymptotics~(\ref{38}) turns out to be of the order of 5\% . To us this is a sufficiently convincing reason to try a similar scheme in the multi-component case. The derivation is given in Appendix \ref{B.b}. The discussion there holds for general $n$ and for general coupling constants $G^\alpha_{\beta\gamma}$. The purpose of this section is to narrow the theory down to the form specific for anharmonic chains.

In the normal mode representation, numerically it is observed that the solution to the mode-coupling equations becomes very quickly diagonal. To analyse the long time asymptotics we therefore rely on the diagonal representation
\begin{equation}\label{40}
  S^{\sharp\phi}_{\alpha\beta}(x,t)=\langle \phi_\alpha(x,t)\phi_\beta(0,0)\rangle\simeq \delta_{\alpha\beta} f_\alpha(x,t)\,.
\end{equation}
Then $f_\alpha(x,0)=\delta(x)$ and the $f_\alpha$'s satisfy
\begin{equation}\label{41}
\partial_t f_\alpha(x,t)= (-c_\alpha \partial_x+D_\alpha \partial^2_x) f_\alpha (x,t) + \int^t_0 ds \int_{\mathbb{R}} dy
  f_\alpha(x-y,t-s) \partial^2_y M_{\alpha\alpha}(y,s)\,,
\end{equation}
$\alpha=-1,0,1$, $D_{\alpha\alpha}=D_\alpha$, with memory kernel
\begin{equation}\label{42}
M_{\alpha\alpha}(x,t)=2 \sum_{\beta,\gamma=0,\pm 1} (G^\alpha_{\beta\gamma})^2 f_\beta(x,t) f_\gamma(x,t)\,.
\end{equation}

For given $p,\beta$, and $V$ the couplings $G^\alpha_{\beta\gamma}$ are listed in Appendix \ref{A.d}. As will be discussed, the long time limit of~(\ref{41}) is dominated by the diagonal coefficients $G^\alpha_{\beta\beta}$. Note that $G^0_{00}=0$ always, also $G^1_{11}= -G^{-1}_{-1-1}$. The case $G^1_{11} \neq 0$ is referred to as ``standard''. A widely studied, but exceptional, case is $p=0$ and a potential which is reflection symmetric with respect to some $y_0$, \textit{i.e.} $V(y - y_0) = V(-y-y_0)$. This class includes the FPU $\beta$ chain with $V(y)=\frac{1}{2}y^2 +\frac{1}{4} \mathsf{b} y^4$, $\mathsf{b}>0$, and the purely quartic potential. For such chains $G^1_{11}=0$. The condition $p=0$ is required. For $p\neq 0$ generically one is back to a standard chain. Both cases will be studied separately.\smallskip\\
(i) \textit{standard case}. Under our assumptions $c>0$. This means that the two sound mode peaks are centered at 
$\pm ct$ and the heat mode peak at $0$. All three peaks are expected to have a width much less than $ct$. But then in~(\ref{42}) the product  $f_\beta(x,t) f_\gamma(x,t) \simeq 0$ for large $t$ in case $\beta\neq \gamma$. Hence the memory kernel can be approximated as
\begin{equation}\label{43}
M_{\alpha\alpha}(x,t)\simeq M^{\mathrm{dg}}_{\alpha}(x,t)=2\sum_{\gamma=0,\pm 1}(G^\alpha_{\gamma\gamma})^2 f_\gamma(x,t)^2\,.
\end{equation}
$f_\alpha$ travels with velocity $c_\alpha$ and in Eq.~(\ref{41}) the product $f_\alpha \partial_y^2M^{\mathrm{dg}}_{\alpha}$ 
will make a significant contribution only if $\gamma = \alpha$.
For $\alpha=0$, one has $G^0_{00}=0$ which requires to investigate
the subleading  terms.  For $\alpha= \pm 1$, as will be explained in Section \ref{sec6}, while the terms with $\gamma \neq \alpha$ are eventually small, still they give rise to long-lived correction terms. 

For the sound peaks, $\sigma=\pm 1$, in leading order one arrives at
\begin{equation}\label{44}
\partial_t f_\sigma(x,t)=(-\sigma c \partial_x+D_\sigma \partial^2_x) f_\sigma(x,t)+ 2(G^\sigma_{\sigma\sigma})^2 \int^t_0 ds \int_{\mathbb{R}} d y
 f_\sigma(x-y,t-s) \partial^2_yf_\sigma(y,s)^2\,,
\end{equation}
which is the one-loop mode-coupling equation for the stochastic Burgers equation. For large $x,t$, its solution with 
initial condition $f_\sigma(x,0) = \delta(x)$ takes the scaling form
\begin{equation}\label{45}
f_\sigma(x,t)\cong (\lambda_\mathrm{s} t)^{-2/3} f_{\mathrm{mc}} \big((\lambda_\mathrm{s} t)^{-2/3}(x-\sigma ct)\big)\,.
\end{equation}
Inserting this scaling form in~(\ref{44}) and following the Fourier transform conventions of Appendix \ref{B.c}, one first finds that
the non-universal scaling coefficient
\begin{equation}\label{46}
\lambda_\mathrm{s} = 2\sqrt{2}|G^\sigma_{\sigma\sigma}|\,,
\end{equation}
in accordance with the KPZ scaling theory. Secondly 
$\hat{f}_{\mathrm{mc}}$ is defined as the solution of fixed point equation
\begin{equation}\label{45a}
\tfrac{2}{3} \hat{f}'_{\mathrm{mc}}(w) = - \pi^2 w \int_0^1 ds \hat{f}_{\mathrm{mc}}((1-s)^{2/3}w) \int_{\mathbb{R}} dq
\hat{f}_{\mathrm{mc}}(s^{2/3}(w-q)) \hat{f}_{\mathrm{mc}}(s^{2/3} q)
\end{equation}
with $w \geq 0$ and $\hat{f}_{\mathrm{mc}}(0) = 1$, $\hat{f}'_{\mathrm{mc}}(0) = 0$.

There is no reason to believe that~(\ref{45}) is the true asymptotic behavior of the sound mode peak. Using an expansion to all orders, it is argued in~\cite{vB12} that the exact sound mode scaling function should be the same as the one of the 
stochastic Burgers equation, hence 
\begin{equation}\label{45b}
S^{\sharp\phi}_{\sigma\sigma} (x,t)\cong (\lambda_\mathrm{s} t)^{-2/3} f_{\mathrm{KPZ}} \big((\lambda_\mathrm{s} t)^{-2/3}(x-\sigma ct)\big)\,.
\end{equation}
As it turns out, $f_{\mathrm{KPZ}}$ differs from $f_{\mathrm{mc}}$ by a few percent only. 

To study the heat mode peak, $f_0$, we switch to Fourier space
\begin{eqnarray}\label{47}
&&\hspace{-30pt}\partial_t \hat{f}_0(k,t)= -D_0 (2\pi k)^2  \hat{f}_0(k,t)\nonumber\\
&&\hspace{0pt} -2\sum_{\sigma=\pm 1} (G^0_{\sigma\sigma})^2(2\pi  k)^2 \int^t_0 ds  \hat{f}_0(k,t-s) 
\int_{\mathbb{R}}  dq  \hat{f}_\sigma(k-q,s)  \hat{f}_\sigma(q,s)\,,
\end{eqnarray}
$ \hat{f}_\sigma(k,0)=1$. For $\hat{f}_\sigma$ one inserts the asymptotic result~(\ref{45}). Then~(\ref{47}) is a linear equation which is solved through Laplace transform, see Appendix \ref{app.C} for details. In our particular case, 
\begin{equation}\label{48}
\hat{f}_0(k,t)\cong \mathrm{e}^{-|k|^{5/3} \lambda_\mathrm{h} t}\,,
\end{equation}
where
\begin{eqnarray}\label{49}
&&\hspace{-15pt}\lambda_\mathrm{h}= \lambda^{-2/3}_\mathrm{s} (G^0_{\sigma\sigma})^2 (4 \pi)^2  \int^\infty_0 dt t^{-2/3} \,\cos (2 \pi ct) \int_{\mathbb{R}}  dx f_{\mathrm{mc}}(x)^2 \nonumber\\
&&\hspace{0pt} =  \lambda^{-2/3}_\mathrm{s} (G^0_{\sigma\sigma})^2 (4 \pi)^2 (2\pi c)^{-1/3}
\tfrac{1}{2}\pi \frac{1}{\Gamma(\tfrac{2}{3})}\frac{1}{ \cos(\tfrac{\pi}{3})} \int_{\mathbb{R}}  dx f_{\mathrm{mc}}(x)^2
\end{eqnarray}
and we used the symmetry $G^0_{\sigma\sigma} = -G^0_{-\sigma-\sigma}$. (\ref{48}) is the Fourier transform of the symmetric $\alpha$-stable distribution with exponent $\alpha=5/3$, also known as Levy distribution. In real space the asymptotics reads, for $|x|\geq (\lambda_\mathrm{h} t)^{3/5}$,
\begin{equation}\label{50}
f_0(x,t)\simeq \pi^{-1} \lambda_\mathrm{h} t |x|^{-8/3}\,.
\end{equation}
$f_\mathrm{mc}$ is a smooth function with rapid decay. On the other hand, $f_0$ has fat tails and its variance is divergent. According to~(\ref{50}), at $x=\sigma ct$ one has $f_0(\sigma ct,t)\cong \pi^{-1} \lambda_\mathrm{h} c^{-8/3} t^{-5/3}$. This explains why there is still coupling between $f_0$ and $f_\sigma$, despite the large spatial separation. In fact, numerically one observes that beyond the sound cone, $x = \pm ct$, the solution decays exponentially fast. As $t$ becomes large the tails of $f_0$ are build up in between the two sound peaks. In the hypothetical case $G^0_{00}\neq 0$, all three modes would have KPZ scaling as in~(\ref{45b}) for $t\to\infty$.\smallskip\\
(ii) \textit{even potential}, $p=0$. The potential satisfies $V(y - y_0) = V(-y-y_0)$ for some $y_0$. As discussed in Appendix \ref{A.5}, the only non-zero couplings are $G^\sigma_{0\sigma'}$, $G^0_{\sigma\sigma}$ and the $R^{-1}$ matrix degenerates to
\begin{equation}\label{50c}
R^{-1}= (Z_{1})^{-1}\begin{pmatrix}
-1& 0 &- 1\\ - c&0&c
\\ 0&-\kappa\partial_\ell p&0
\end{pmatrix}\,.
\end{equation}
Hence the $\ell$-$\ell$  correlations are $(Z_1)^{-2}(f_{-1} + f_{1})$ and the $\mathsf{u}$-$\mathsf{u}$ correlations  $(Z_1)^{-2}c^2(f_{-1} + f_{1})$, no heat mode peak, while the
  $\mathsf{e}$-$\mathsf{e}$ correlations are $(Z_1)^{-2}(\kappa\partial_\ell p)^2f_{0}$, no sound mode peaks. The mode-coupling equation simplifies to 
\begin{eqnarray}\label{50a}
&&\hspace{-40pt}\partial_t f_\sigma(x,t)= (-\sigma c \partial_x +D_\sigma \partial^2_x)  f_\sigma(x,t)\nonumber\\
&&\hspace{20pt}+\sum_{\sigma'=\pm 1} 2(G^\sigma_{0\sigma'})^2 \int^t_0 ds \int_{\mathbb{R}}  dy  (\partial^2_y
f_\sigma(x-y,t-s))  f_0(y,s)f_{\sigma'}(y,s)\,.
\end{eqnarray}
Since the product $ f_0f_{\sigma'}$ is very small, the diffusive term will dominate for long times and yields
\begin{equation}\label{51}
f_\sigma(x,t)=\frac{1}{\sqrt{4\pi D_\sigma t}}\mathrm{e}^{-(x-\sigma ct)^2/4 D_\sigma t}
\end{equation}
asymptotically. 
The sound mode is diffusive. As for the one-component case in principle there could be logarithmic corrections coming from the cubic term in the expansion of the Euler currents. If finite, $D_\sigma$  would be determined by the Green-Kubo formula, \textit{i.e.}  by the time-integral of the current-current correlation for mode $\sigma$.

$f_0$ is governed by
\begin{eqnarray}\label{50b}
&&\hspace{-30pt}\partial_t f_0(x,t)= D_0 \partial^2_x  f_0(x,t)\nonumber\\
&&\hspace{30pt}+\sum_{\sigma=\pm 1} 2(G^0_{\sigma\sigma})^2 \int^t_0 ds \int_{\mathbb{R}}  dy  (\partial^2_y
f_\sigma(x-y,t-s))  f_{\sigma}(y,s)^2\,.
\end{eqnarray}
We can now proceed as under (i). Only now $f_\sigma$ is given by~(\ref{51}) rather than~(\ref{45}). From Appendix C we conclude
\begin{equation}\label{52}
\hat{f}_0(k,t)=\mathrm{e}^{-|k|^{3/2} \lambda_\mathrm{h}t}
\end{equation}
with
\begin{equation}\label{53}
\lambda_\mathrm{h}= (D_\sigma)^{-1/2} (G^0_{\sigma\sigma})^2 (4\pi)^2(2\pi c)^{-1/2} \int^\infty_0 dt \,t^{-1/2} \cos (t)
(2\sqrt{\pi})^{-1}\,,
\end{equation}
the integral being equal to $\sqrt{\pi/2}$. The $\tfrac{3}{2}$-Levy distribution is broader than the $\tfrac{5}{3}$-Levy
distribution.
Thus a diffusive sound mode seems to produce a broader heat mode, which is somewhat counter-intuitive.  


\section{Total current correlations}\label{sec5}
\setcounter{equation}{0}

Besides the space-time correlator, a further central object are the current correlations, in particular the total current correlations. There is a link through the conservation laws, but it is more instructive to discuss the current correlations on their own. For the anharmonic chain the space-time current correlations are defined through the equilibrium average
\begin{equation}\label{55}
C^\mathcal{J}_{\alpha\beta}(j,t)= \langle \mathcal{J}_{\alpha}(j,t) \mathcal{J}_{\beta}(0,0)\rangle_{p,\beta} - \langle \mathcal{J}_{\alpha}(0,0)\rangle_{p,\beta}\langle \mathcal{J}_{\beta}(0,0)\rangle_{p,\beta}\,,
\end{equation}
where $\vec{\mathcal{J}}(j,t)=\big(-p_j(t),-V'(r_{j-1}(t)),-p_j(t) V'(r_{j-1}(t))\big)$. The corresponding objects can also be defined for nonlinear fluctuation hydrodynamics. According to~(\ref{34}) the normal mode currents are
\begin{equation}\label{56}
\mathcal{J}^{\phi}_\alpha(x,t)= c_\alpha \phi_\alpha(x,t)+\langle\vec{\phi}(x,t), G^\alpha \vec{\phi}(x,t)\rangle -\partial_x (D\vec{\phi})_\alpha(x,t) +(B\vec{\xi})_\alpha(x,t)
\end{equation}
and the current-current correlation functions are defined by
\begin{equation}\label{57}
C^{\sharp\phi}_{\alpha\beta} (x,t)=\langle \mathcal{J}^\phi_\alpha (x,t) \mathcal{J}^{\phi}_\beta (0,0)\rangle-\langle \mathcal{J}^\phi_\alpha (0,0)\rangle\langle \mathcal{J}^\phi_\beta (0,0)\rangle\,,
\end{equation}
average in the stationary mean zero $\phi(x,t)$ process. As in~(\ref{26a}), to connect to the microscopic correlation $C^\mathcal{J}$ one still has to undo the transformation to normal modes, resulting in the central conjecture
\begin{equation}\label{58}
C^\mathcal{J}(j,t)\simeq (R^{-1} C^{\sharp\phi} (R^{-1})^\mathrm{T})(x,t)\,,
\end{equation}
supposed to be valid on a mesoscopic scale. As before, $\epsilon j = \lfloor x\rfloor_\epsilon$ with sufficiently small lattice spacing $\epsilon$.

In molecular dynamics, mostly one studies the total current correlation defined by
\begin{equation}\label{59}
C^{\mathrm{tot}} (t)=\sum_{j \in \mathbb{Z}} C^\mathcal{J}\!(j,t)\,,\quad C^{\sharp\phi,\mathrm{tot}} (t)=
\int_{\mathbb{R}} dx  C^{\sharp\phi}(x,t)\,.
\end{equation}
This means to consider in Fourier space the limit $k\to 0$ at fixed $t$.

As a general fact, current correlations are linked to the memory kernel. This holds also for nonlinear fluctuating hydrodynamics. As explained in Appendix \ref{B.b}, within the one-loop approximation,
\begin{eqnarray}\label{60}
&&\hspace{-20pt} C^{\sharp\phi,\mathrm{tot}}_{\alpha\alpha'} (t)= \int_{\mathbb{R}} dx M_{\alpha\alpha'} (x,t)
= 2 \int _{\mathbb{R}}dx \mathrm{tr} [S^{\sharp\phi} (x,t)^\mathrm{T} G^\alpha S^{\sharp\phi} (x,t) G^{\alpha'}]\\
&&\hspace{10pt}= 2\int_{\mathbb{R}} dk \mathrm{tr} [\hat{S}^{\sharp\phi} (-k,t)^\mathrm{T} G^\alpha \hat{S}^{\sharp\phi} (k,t) G^{\alpha'}]
\simeq \sum_{\gamma=0,\pm 1} 2 G^\alpha_{\gamma\gamma} G^{\alpha'}_{\gamma\gamma} \int_{\mathbb{R}} dx f_\gamma (x,t)^2 \,,
\nonumber
\end{eqnarray}
where in the last step we used the diagonal approximation for $S^{\sharp\phi}$ and the small overlap between peaks moving with distinct velocities. We now narrow down to the standard and  $V$ even, $p=0$ cases.\smallskip\\
(i) \textit{standard case}. Using~(\ref{45}) and~(\ref{48}), one arrives at the leading long time behavior
\begin{eqnarray}\label{61}
&&\hspace{-10pt}C^{\sharp\phi,\mathrm{tot}}_{\alpha\alpha'} (t)= 4 G^\alpha_{11} G^{\alpha'}_{11} (\lambda_\mathrm{s} t)^{-2/3} \int _{\mathbb{R}} dx f_{\mathrm{KPZ}} (x)^2\nonumber\\
&&\hspace{54pt} +\, 2G^\alpha_{00} G^{\alpha'}_{00} (\lambda _{\mathrm{h}} t)^{-3/5} \int _{\mathbb{R}} dk \mathrm{e}^{-2|k|^{5/3}}\,.
\end{eqnarray}
In particular for the sound mode current, $\alpha=\alpha'=\pm 1$, the leading decay is $t^{-3/5}$, while for the heat mode current, $\alpha=\alpha'=0$, the second term does not contribute  since $G^0_{00}=0$ and the asymptotic decay is $t^{-2/3}$. To reiterate, predicted is not only the power law but also the non-universal prefactor. One first has to transform 
Eq.~(\ref{55}) to normal mode currents and sum over $j$. Then in the limit $t \to \infty$, say, the (11)-matrix element decays as 
$t^{-3/5}$ with prefactor as stated in~(\ref{61}). \smallskip\\
(ii) \textit{even $V$, $p=0$}. The only nonintegrably decaying current correlation is the heat current
\begin{equation}\label{63}
C^{\sharp\phi,\mathrm{tot}}_{00} (t) \simeq 4 (G^0_{11})^2(2\pi D_{+} 2t)^{-1/2}\,.
\end{equation}
The sound current must be determined by terms not included in~(\ref{60}), which is consistent with an integrable decay of the sound mode current, baring possible logarithmic corrections from third order terms.

\section{Corrections to scaling}\label{sec6}
\setcounter{equation}{0}

For the long time asymptotics we argued, see~(\ref{45}) and~(\ref{48}),
\begin{eqnarray}
&&\hspace{-10pt} \hat{f}_\sigma(k,t)\cong \hat{f}_{\mathrm{mc}}\big((\lambda_\mathrm{s} t)^{2/3}k\big) \mathrm{e}^{-\mathrm{i} 2 \pi\sigma ct k}\,,\\\label{6.10}
&&\hspace{-10pt} \hat{f}_0(k,t)\cong \hat{f}_{\mathrm{h}} \big((\lambda_\mathrm{h} t)^{3/5}k\big)\,,\quad \hat{f}_{\mathrm{h}}(k)= \mathrm{e}^{-|k|^{5/3}}\,.
\end{eqnarray}
In numerical solutions of the mode-coupling equations it is observed that for $k=\mathcal{O}(t^{-2/3})$ the sound mode $\hat{f}_1(k,t)$ converges fairly quickly to a definite shape, which however is still slowly evolving towards $\hat{f}_\mathrm{mc}$. On the other side on the scale $k=\mathcal{O}(t^{-3/5})$ the heat mode $\hat{f}_0(k,t)$ converges rapidly to $\hat{f}_{\mathrm{h}}$. In fact, outside the sound cone, $\{|x|\geq ct\}$, the correlator is exponentially small, while $f_{\mathrm{h}}(x)$ is being build up  in the interior. Of course, the dynamics depends on all coupling constants. To have a very rough idea: for $c=1$, $|G^\sigma_{\alpha\alpha}|=1$, $|G^{0}_{\sigma\sigma}|=1$, all other couplings zero while respecting the symmetries of the model, for times $t\gtrsim 100$ only the slow convergence of the sound modes persists. One could try to build a two-scale theory. But, at the moment, a simple order of magnitude result has to suffice.

We use the diagonal approximation discussed in Section 4. For the heat mode $G^0_{00}=0$ and $G^0_{
\sigma\sigma}$ is used already for the asymptotics~(\ref{6.10}). Hence the corrections should be small, consistent with the numerical findings. For the sound mode, in addition to $G^\sigma_{\sigma\sigma}$, we now include the effects coming from $G^\sigma_{00}$, $G^\sigma_{-\sigma-\sigma}$. Let us consider mode 1 in the frame moving with $ct$, $\hat{f}_1(k,t)=\mathrm{e}^{-\mathrm{i}2 \pi ckt} \hat{h}(k,t)$. Then $\hat{h}$ satisfies
\begin{equation}\label{6.3}
\partial_t \hat{h}(k,t)=-(2\pi k)^2 \int^t_0 ds \hat{h}(k,t-s) \hat{M}(k,s)
\end{equation}
with the memory kernel
 \begin{eqnarray}\label{6.4}
&&\hspace{-20pt} \hat{M}(k,s)= 2(G^1_{11})^2 \int_{\mathbb{R}} dq \hat{h}(k-q,s) \hat{h}(q,s)\\
&&\hspace{-20pt} +\,\mathrm{e}^{\mathrm{i}2\pi cks} \Big(2(G^1_{00})^2
\int _{\mathbb{R}}dq \hat{f}_0 (k-q,s) \hat{f}_0(q,s) +2(G^1_{-1-1})^2 \int_{\mathbb{R}} dq \hat{f}_{-1}(k-q,s) \hat{f}_{-1}(q,s)\Big)\,.\nonumber
\end{eqnarray}
$\hat{f}_{-1}(k,s)$ carries the phase factor $\mathrm{e}^{\mathrm{i} 2 \pi cks}$. Thus to have a contribution to the integral one should take $k=\mathcal{O}(s^{-1})$, whereas the variation of $\hat{h}$ is on the scale $k=\mathcal{O}(t^{-2/3})$. This suggest that in the second and third term we may set $k=0$. Assuming already the scaling form, 
\begin{eqnarray}\label{6.5}
&&\hspace{-20pt}\int_{\mathbb{R}} dq \hat{f}_0 (-q,s) \hat{f}_0(q,s)= \int_{\mathbb{R}} dx |f_0(x,s)|^2 =s^{-3/5} a_0(s)\,,
\nonumber\\
&&\hspace{-20pt}\int_{\mathbb{R}} dq \hat{f}_{-1} (-q,s) \hat{f}_{-1}(q,s)= \int_{\mathbb{R}} dx |f_{-1}(x,s)|^2 =s^{-2/3} a_{-1}(s)\,.
\end{eqnarray}
$a_0$, $a_{-1}$ are insensitive to fine details of $f_0$, resp. $f_{-1}$. They vary slowly and $|a_0|$, $|a_{-1}|$ are bounded away from 0. Hence the memory kernel can be written in approximation as 
 \begin{eqnarray}\label{6.7}
&&\hspace{-20pt} \hat{M}(k,s)\simeq 2(G^1_{11})^2 \int _{\mathbb{R}}dq \hat{h}(k-q,s) \hat{h}(q,s) + 2(G^1_{00})^2 a_0(s) (1+s)^{-3/5}\mathrm{e}^{\mathrm{i} 2 \pi cks}\nonumber\\
&&\hspace{32pt} +2 (G^1_{-1-1})^2 a_{-1}(s) (1+s)^{-2/3} \mathrm{e}^{\mathrm{i} 2 \pi 2cks}\,.
\end{eqnarray}

The relative size of the terms can be estimated by using for $\hat{h}$ its asymptotic scaling form, $\hat{h}(k,t)=\hat{f}_{\mathrm{mc}}\big((\lambda_\mathrm{s} t)^{2/3}k\big)$. Substituting $s$ by $st$ and $k$ by $wt^{-2/3}$ the first summand of~(\ref{6.3}) together with~(\ref{6.7}) becomes
\begin{eqnarray}\label{6.8}
&&\hspace{-30pt}-t^{-1} 2(G^1_{11})^2(2 \pi w)^2 \int^1_0 ds \hat{f}_{\mathrm{mc}} ((\lambda_\mathrm{s} (1-s))^{2/3}w) \nonumber\\
&&\hspace{40pt} \times \int_{\mathbb{R}}dq \hat{f}_{\mathrm{mc}} ((\lambda_\mathrm{s} s)^{2/3}(w-q)) \hat{f}_{\mathrm{mc}} ((\lambda_\mathrm{s} s)^{2/3}q)\,.
\end{eqnarray}
For the two other terms one has to choose $k=w t^{-2/3}$ because of the scaling of $\hat{h}$. The oscillatory factor $\mathrm{e}^{\mathrm{i} 2 \pi cks}$ forces the substitution of $s$ by $st^{2/3}$. Then the second and third summand of~(\ref{6.3}) together with~(\ref{6.7}) read
\begin{eqnarray}\label{6.9}
&&\hspace{-20pt} -t^{-1} (2 \pi w)^2 t^{-1/15} 2(G^1_{00})^2 a_0(\infty) \hat{f}_\mathrm{mc} (\lambda^{2/3}_\mathrm{s} w)\int^\infty_0 ds s^{-3/5} \mathrm{e}^{\mathrm{i} 2 \pi cws} \,,
\nonumber\\
&&\hspace{-20pt} -t^{-1} (2\pi w)^2 t^{-1/9} 2(G^1_{-1-1})^2 a_{-1}(\infty) \hat{f}_\mathrm{mc} (\lambda^{2/3}_\mathrm{s} w)\int^\infty_0 ds s^{-2/3} \mathrm{e}^{\mathrm{i} 2 \pi 2cws} \,.
\end{eqnarray}

In conclusion, while the correction terms for the sound modes vanish eventually, they decay very slowly. One has to be prepared that the scaling of the sound mode peaks is seen only for very long times. Mode-coupling provides approximate,
but physically interesting information on the dynamics of correlations at intermediate time scales.
\section{A guide to the literature with comments }\label{sec7}
\setcounter{equation}{0}
In focus are the equilibrium time correlations of  hamiltonian particle systems in one spatial dimension, either classical or quantum. A first important distinction is between  non-integrable and integrable systems. We encountered already examples of the latter, as the harmonic chain and the chain with hard-core equal mass particles. More examples are the Toda chain, classical and quantum, the 
Lieb-Liniger model ($\delta$-Bose gas) and a few more. Nonlinear fluctuating hydrodynamics deals only with non-integrable systems at non-zero temperature. At $T=0$ classical systems degenerate, while quantum systems still have a rich phase diagram and even richer dynamical properties, very much in focus of current research.

In our note we have studied the equilibrium time correlations of the conserved fields. Physically this means that the system is prepared in thermal equilibrium and probed by a small localized perturbation which spreads out in the course of time, as dictated by the transport of elongation, momentum, and energy. A related but conceptually distinct way to probe transport is to couple the system at both border points to suitable reservoirs and to monitor the steady state transmission, in most of the literature the transport of energy. On this topic there is a huge literature, for which we have to refer to the reviews
\cite{LeLi03,Dh08}. There is no difficulty to add to~(\ref{34}) boundary terms which model the injection and removal 
of a conserved quantity to/from the system. But currently we have no tools how to handle the resulting 
stochastic field theory.\bigskip \\
\textbf{7.1 Equilibrium time correlations} \medskip\\
In one dimension, while static correlations decay rapidly, the correlations in time have anomalously slow decay. 
Such long-time tails were discovered by Alder and Wainwright \cite{AlWa70} in the context of velocity autocorrelations for hard disks and hard spheres fluids. In a pioneering work, Ernst, Hauge, and van Leeuwen~\cite{ErHa76} 
developed a mode-coupling theory for general dimension $d$. Extending naively to $d=1$, a decay as $t^{-1/2}$
is predicted, indicating that the interaction of modes has to be reanalyzed. Such a  program has been accomplished only
recently by van Beijeren \cite{vB12}.
Forster, Nelson, and Stephen~\cite{FoNe77} consider the incompressible Navier-Stokes equation with
random velocity currents added. For $d=1$ this equation looses its meaning. Instead the random Burgers equation
(\ref{36}) is proposed and studied, already noting the dynamical exponent $3/2$.

For quantum systems Andreev~\cite{An80} discovered the $k^{3/2}$
scaling of the sound peak. His derivation has been  improved and expanded in~\cite{Sa80,PuZw05, KuLa12,ArBo13}.

There is great interest in dynamical properties of one-dimensional quantum systems. But these investigations  do not seem to reach the equilibrium time correlations. Also their DMRG type simulation is demanding.
Hence we narrow down to classical systems. \medskip\\
\textbf{i) Molecular dynamics simulations}\medskip\\
\textit{FPU chains}.  The FPU potential is $V(x) = \tfrac{1}{2}x^2 +  \tfrac{1}{3}\mathsf{a}x^3 +  \tfrac{1}{4}\mathsf{b}x^4$. (To avoid double meaning the historial $\alpha$ is replaced by $\mathsf{a}$ and $\beta$ is replaced by 
$\mathsf{b}$). A widely studied case is $\mathsf{a} = 0$,  $\mathsf{b} = 1$, and the purely quartic model, at pressure $p=0$.
 This is the even potential case which has very special features,
 as first pointed out and observed numerically in \cite{LeLiPo03} and also discussed in the main text.
  In~\cite{LeDa05} total energy and heat current correlations are studied, while in~\cite{Zh06,Zh13} the spatially resolved correlations are reported.
According to 
(\ref{63}) the asymptotic decay should be $t^{-1/2}$, assuming that the sound modes are diffusive. 
The most recent  simulations~\cite{WaWa11} measure the energy current correlation for $N$ up to $2 \times 10^4$. However the predicted power law is not observed convincingly.
For the symmetric potential a decay of the heat current as $t^{-2/3}$ is reported in~\cite{ChZh12}.   In~\cite{ZaDe13}
the energy-energy correlations are studied in case $\mathsf{a} = 0$, $\mathsf{b} = 1$, $p = 0$, and $\mathsf{e} = 1$
for a chain of 16.000 particles. For the heat mode peak Levy 5/3 is reported, as compared to Levy 3/2 from the theory,
while the diffusive spreading of the sound mode peaks is not directly investigated.

The asymmetric case has been studied in~\cite{LeDa07,WaWa11}. The predicted $t^{-2/3}$ decay is consistent
with the numerical findings. Exponential decay is reported in ~\cite{ChZh12}. 
In~\cite{LeDa07} it is argued that the sound mode and the heat mode currents have the same asymptotic decay. In frequency
space this would mean that the ratio, called Prandtl number,
\begin{equation}\label{62}
\hat{C}^{\sharp\phi,\mathrm{tot}}_{00} (\omega)/\hat{C}^{\sharp\phi,\mathrm{tot}}_{11} (\omega)
\end{equation}
has a non-zero limit as $\omega \to 0$. Mode-coupling arrives at a different conclusion.

As a rule, for even potentials at $p=0$ a definite power law decay seems to be more difficult to establish than 
for asymmetric potentials.
\medskip\\
\textit{Chains with other smooth potentials.}~\cite{ChZh12} a Lennard-Jones type potential is investigated. In~\cite{ZhZh12} the potential $V(x) = \tfrac{1}{2}(x+ r)^2 + \mathrm{e}^{-rx}$ is studied. Diffusive transport is claimed. Chains with transverse motions are simulated  
in~\cite{WaLi04}. \medskip\\
\textit{Hard-point particles}.
Particles have a point hard core and thereby maintain their order. As a chain, this corresponds to the potential $V(x) = 0$ for $x>0$ and $V(x) = \infty$ for $x\leq 0$. Then $p(\ell,\mathsf{e}) = 2 \mathsf{e}/ \ell$. Nonlinear fluctuating hydrodynamics makes nontrivial predictions, but the true dynamical behavior is that of an ideal gas. This example underlines again the requirement for the dynamics to be sufficiently chaotic. To improve the model, it is assumed that the particle mass depends on the index $j $.
The standard choice  is $m_j$ to have period 2, \textit{i.e.} the masses are alternating. Then only particles of unequal mass collide, which provides a mechanism for presumably sufficiently strong dynamical mixing. 
The unit cell contains now two particles, which is not directly covered by the main text. Since there are many numerical studies of the hard-point gas, we explain the necessary modifications in Appendix \ref{A.h}. Because of its simple
thermodynamics the hard-point gas serves also as a nice illustration of the general theory.
One variant of the hard-point gas is to impose in addition the constraint of a maximal distance, say $a$. Once neighboring particles reach that distance, then they are reflected inwards. This corresponds to $V(x) = 0 $ for $0< x < a$ and $V(x) = \infty$ otherwise.

First simulation results were obtained by Grassberger et al.~\cite{GrNa02} for $a = \infty$. They confirmed the  $t^{-2/3}$ decay
of the total energy current, which led them to first conjecture a tentative connection to KPZ. 
In  ~\cite{CiDe05} the spreading of energy is reported with an extremely accurate scaling plot of the $\tfrac{5}{3}$-Levy distribution. The $\tfrac{5}{3}$-Levy is guessed on the basis of the numerical data. While this agrees with the prediction of mode-coupling, not exactly the same correlation is measured. A more recent numerical study of the same quantity is~\cite{ZaDe13}.
The case $a < \infty$ is simulated in ~\cite{DeDe07} and for the heat mode good agreement with $\tfrac{5}{3}$-Levy
is found, which is also an indirect evidence for the scaling of sound modes. $p=0$ is identified as a special point in parameter space, with more extensive confirmations in \cite{Po11}. Since the potential is symmetric relative to $y_0 = a/2$, from the perspective of nonlinear fluctuating hydrodynamics this special point is in the same class as the FPU chain with 
even potential at $p=0$. The numerical simulation very convincingly reports a diffusive broadening of the heat peak,
at variance with the discussion in Section \ref{sec4}. For more recent simulations we refer to~\cite{Zh13}.
In~\cite{DeNa03} the deterministic collision are replaced by random collision. The $t^{2/3}$ spreading of the sound mode is clearly observed.\medskip\\
\textit{Fluids.} In the literature, the hard-point particles are often referred to as fluid. While one might take this point of view,
hard-point particles are a particular chain and covered by our methods. Also Lennard-Jones fluid mostly means a chain, for which $V$ is a Lennard-Jones type potential.\medskip\\
\textbf{ii) Theoretical approaches}\medskip\\
\textit{Hydrodynamic limit for anharmonic chains}.
The equivalence of ensembles and the derivation of the Euler equations are covered extensively in the pre-version of  the monograph 
\cite{Olla12}. To have sufficient time mixing the authors add stochastic collisions to the hamiltonian evolution.
\medskip\\
\textit{Mode-coupling equations}.
Mode-coupling somewhat vaguely refers to have an approximate but closed equation for the correlator. Generically, the mode-coupling equations are nonlinear and of memory type. One arrives at them by suitable closure assumptions. A more systematic approach is a Mori-Zwanzig projection onto to the conserved fields. For stochastic field theories, like~(\ref{34}),
one can use the division between the Gaussian and nonlinear part to expand in the nonlinearity. A partial resummation of the diagrams leads to a mode coupling equation. A recent account is~\cite{vB11}.

For anharmonic chains early attempts are~\cite{Schir,Le98}.  In these works the positional correlations $\sum_j \exp[-\mathrm{i} k q_j(t)]$ are studied. Later on the scheme has been extended to energy correlations ~\cite{DeLe06,DeDe08}. As argued in~\cite{vB12}, and in more detail here, 
for a systematic theory one has to consider the full $3\times 3$ time-correlator.

For one-dimensional fluids, non-reducible to chains, the starting equations are~(\ref{D.2}), (\ref{D.3}). The resulting transformation to normal modes is very similar. As main difference, the pressure and the internal energy are not as explicitly available  as for chains. Currently,~\cite{vB12,vB11} provide  the most detailed coverage. The asymptotic results are in agreement with our approach, including the values of the non-universal coefficients.  \medskip\\
\textit {Dynamical scaling and RG analysis}.
In the context of FPU chains a RG type analysis, combined with hydrodynamics, has been attempted in~\cite{LeDa05}. Dynamical scaling based on the full nonlinear Navier-Stokes equations with added white noise random currents is carried out in~\cite{NaRa02}. Eventually such stochastic equations will have to be treated by RG methods.   For the comparison with MD simulations information beyond the fixed point theory is of interest, which is more readily provided in the one-loop approximation.
\bigskip\\
\textbf{7.2 Noisy Burgers equation} \medskip\\
The noisy Burgers equation reads 
 \begin{equation}\label{7.3a}
 \partial_t u +\partial_x( \tfrac{1}{2}u^2 - \tfrac{1}{2}\partial_x u + \xi) = 0
 \end{equation}
 with $\xi(x,t)$ standard space-time white noise, which corresponds to $n = 1$ in Eq. (\ref{34}).
 Mostly one studies the integrated form, $u = \partial_x h$, and arrives at the one-dimensional KPZ equation 
 \begin{equation}\label{7.3}
 \partial_t h = - \tfrac{1}{2}(\partial_x h)^2 + \tfrac{1}{2}\partial_x^2h - \xi\,.
 \end{equation}
 The invariant
measure for (\ref{7.3a})  is mean zero white noise with covariance
 \begin{equation}\label{7.4}
\langle u(x,t) u(x',t) \rangle = \delta(x-x')\,.
 \end{equation}
Equation~(\ref{7.3})  is singular, since the solution of the linear part is not differentiable in $x$. Still ~(\ref{7.3}) can be mathematically defined through the Cole-Hopf transformation $Z= \mathrm{e}^h$,
 where $Z$ then satisfies the well controlled linear stochastic heat equation, 
 \begin{equation}\label{7.5}
 \partial_t Z = \tfrac{1}{2}\partial_x^2Z + \xi Z\,,
 \end{equation}
  see~\cite{Be97,AmCo11}. In ~\cite{Ha13} a more sophisticated, and more general, RG construction is presented.  The exact two-point function $\langle u(x,t)u(0,0)\rangle$ for the stationary 
  process is computed by using the replica method~\cite{ImSa12}.
   The scaling limit of the replica solution yields then $f_{\mathrm{KPZ}}$.
  As a supporting approach, discrete models have been studied, like the PNG model~\cite{PrSp00,PrSp04} and the TASEP 
 ~\cite{FeSp06}. For these models it is proved that the scaling limit of the covariance of the stationary process yields 
$f_{\mathrm{KPZ}}$.

The validity of nonlinear fluctuating hydrodynamics has been tested also for two specific non-equilibrium initial
conditions, (i) the flat case $h(x,0)=0$ and (ii) the sharp wedge $h(x,0) = -\epsilon^{-1}|x|$ in the limit
$\epsilon \to 0$. Corresponding initial conditions can be constructed. \textit{E.g.}, for a $ 0,1$ lattice gas
(i) is modeled by the alternating initial condition $...01010...$ and (ii) by the step $...111000...\,$.
One compares the fluctuations in the lattice gas with the one obtained from nonlinear fluctuating hydrodynamics
(\ref{7.3}),~(\ref{7.4}). 
Asymptotic agreement is established. There has been a lot of activities from the probabilistic side, see
\cite{Co12} for a summary up to 2011. 
\bigskip\\
\textbf{7.3 Coupled KPZ equations}\medskip\\
In the early 90ies Erta\c{s} and Kardar~\cite{EK92,EK93,Kar98} studied the dynamic roughening of directed lines, as for example dislocation, polymer, or vortex lines, and for that purpose used a model  consisting of two coupled one-dimensional KPZ equations. Prominent further examples, employing the same type of Langevin equations, are sedimenting colloidal suspensions~\cite{LRFB98} and crystals~\cite{RS97}, stochastic lattice gases~\cite{DBBR01}, and magnetohydrodynamics~\cite{Yan97,FD98,BBR99}.  The Langevin equation has the generic structure
\begin{equation}\label{7.1a}
\partial_t h_\alpha = -c_\alpha  \partial_x h_\alpha - \sum_{\beta,\gamma= 1}^n G^\alpha_{\beta\gamma}( \partial_x h_\beta )(
 \partial_x h_\gamma) +  \sum_{\beta= 1}^n  D_{\alpha\beta} \partial_x^2  h_\beta  -
  \sum_{\beta= 1}^n  B_{\alpha\beta}\xi_\beta\,.
\end{equation}
$\alpha$ labels the fields, $\alpha = 1,\ldots,n$. The components $h_\alpha(x,t)$ have varying interpretations depending on the physical context. But obviously, setting $\phi_\alpha = \partial_x h_\alpha$, Eq. (\ref{7.1a}) is identical to (\ref{34}). In~\cite{EK92} the case $n=2$ is studied with the special feature that $c_1 = 0= c_2$. One can no longer use the decoupling argument. Numerical integration yields the same scaling exponent as for $n=1$ but with scaling functions different from $f_{\mathrm{mc}}$. The full parameter space still needs to be explored. In~\cite{DBBR01} 
a moving interface with internal structure is investigated, for which the coupling parameters can be computed and $c_1 \neq c_2$.  The KPZ exponent is validated.\bigskip\\
\textbf{7.4 Interacting stochastic particle systems in one dimension} \medskip\\
``Stochastic particle system'' is a generic name for many degrees of freedom, whose motion is governed by a Markov process, either jump or diffusion, with local interactions. In our context, the standard example are particles of several types evolving in time by random hops on the lattice $\mathbb{Z}$. The jump rates depend on the local environment of the current hop.
Since the number of particles of each type is conserved, by the same argument as given in the main text the dynamics of the fluctuation fields is governed by the stochastic field theory~(\ref{34}). For a lattice gas satisfying detailed balance, the macroscopic currents vanish. To have interesting fluctuation behavior, detailed balance has to be broken (non-reversible dynamics in the probabilistic jargon). By construction, stochastic lattice gases have optimal mixing, only constrained by local conservation laws, and they can be simulated through Monte Carlo. Thus they serve as fine testing ground for (\ref{34}). An early example is~\cite{DBBR01}. In this contribution $n = 2$ and at every site the admissible states are $0,1$. The dynamics are
nearest neighbor hops such that the particle number on the odd and even sublattices are conserved. 
Numerically the 2/3 exponent is observed. Another more recent model is named after Arndt, Heinzel, and 
Rittenberg~\cite{AHR99}. The local states are $0$ and $\pm1$ for the conserved  two types of particles. The steady state is computed
via matrix product ansatz. Hence all coefficients are known analytically. For the normal modes one finds $c_1 \neq c_2$ and also $G^{1}_{11} \neq 0$, $G^{2}_{22} \neq 0$. However the subleading coefficients 
$G^{1}_{22} = 0 = G^{2}_{11}$. In the Monte Carlo simulations~\cite{FeSS13} one observes a rapid relaxation to $f_{\mathrm{KPZ}}$ for each mode, in complete agreement with the theory. Of course, there are other two-component stochastic lattice  gases which have tunable subleading coefficients.

To have models closer to anharmonic chains, one maintains the Hamiltonian time evolution, but adds stochastic collisions between neighboring particles.  They could be merely random exchanges of the two momenta at neighboring sites. One could also add a diffusion in the momentum space for three neighboring particles such that their total momentum and kinetic energy is conserved~\cite{BaBe06,Ol07,Gerschenfeld12}. Modifying even further an
interesting model has been proposed recently~\cite{BeSt12}. The random field is specified by
$\{y_j,j\in\mathbb{Z}\}$ with $y_j \in \mathbb{R}$. The deterministic part of the evolution is governed by
\begin{equation}\label{7.7}
\frac{d}{dt} y_j = V'(y_{j+1}) - V'(y_{j-1})\,.
 \end{equation}
In addition there are random exchanges $\ldots,y_j,y_{j+1},\ldots$ to $\ldots,y_{j+1},y_j,\ldots$ independently at each bond with rate 1. The conserved fields are $y_j$ and $V(y_j)$. The dynamics is non-reversible. The invariant measures are identical to the $\{r_j\}$-part 
of the anharmonic chain. The canonical parameters are $p,\beta$, as before, conjugate to the elongation
$\ell$ and internal potential energy $\mathsf{e}$, 
\begin{equation}\label{7.8}
\ell = \langle y_j\rangle_{p,\beta}\,,\quad\mathsf{e} = \langle V(y_j) \rangle_{p,\beta}\,.
 \end{equation}
There are no momenta. The Euler equations read
 \begin{equation}\label{7.9}
\partial_t \ell + 2\partial_x p =0\,,\quad \partial_t \mathsf{e} - \partial_x p^2 =0
 \end{equation}
with $p = p(\ell,\mathsf{e})$.
Following the standard route one obtains the mode velocities $c_1 =0$, $c_2 = 2(-p\partial_\mathsf{e} p +
\partial_\ell p)$ and the $G$-couplings $G^1_{11} =0$, $G^1_{12} =0$, $G^1_{22} \neq 0$, while $G^2_{\beta\gamma}$
is generically different from 0. Thus the signature is the same as for anharmonic chains with one sound mode missing.
The first mode is predicted to be $\tfrac{5}{3}$-Levy and the second mode to be KPZ.

The harmonic potential, $V(y) = \tfrac{1}{2}y^2$, is degenerate, since $p(\ell,\mathsf{e}) = -\ell$. Then $G^2 = 0$
and on the basis of (\ref{60}) the total (22)-current correlation is conjectured to have a decay as $t^{-1/2}$, in agreement with the proof in \cite{BeSt12}.
In~\cite{BeGo13} the Kac-vanMoerbecke potential $V(y) = e^{-y} -1 +y$ is studied. Then $p(\ell,\mathsf{e}) = \mathsf{e} - \ell$
and the only nonzero $G$-coefficients are
$G^1_{22}$ and $G^2_{22}$. The (11)- and (12)-component of the total current correlation matrix are proved to have
an integrable decay, while the (22)-component  satisfies a lower bound as $const.\, t^{-3/4}$~\cite{BeGo13}. Inserting 
in~(\ref{60})
the $G^{\alpha}$ matrices for the Kac-vanMoerbecke potential  and transforming back to the currents of the physical fields, one obtains the proven signature for the total current correlation matrix with the (22)-component decaying as $t^{-2/3}$. These results I regard as a further indirect confirmation of the validity
of nonlinear fluctuating hydrodynamics as applied to systems of conserved fields.
\noindent\\\\
\textbf{Acknowledgements}. The topic of my article reflects longstanding interests. The actual study was triggered, in fact,  
by two workshops in the fall 2012 on transport in one-dimensional systems, one at the ICTP Trieste,
organized by A. Dhar, M.N. Kiselev, Y.A. Kosevich, R. Livi, and one at BIRS, Banff, organized by J.L. Lebowitz, S. Olla, G. Stoltz, for both of which I am most grateful. I thank H. van Beijeren for sharing his insights
on mode-coupling theory, S. Olla for emphasizing the hydrodynamics of anharmonic chains, J. Krug for pointing at the early literature on coupled KPZ equations, P. Ferrari, C. Mendl, T. Sasamoto for constant help and encouragement, and C. Bernardin, S. Lepri,
A. Politi, H. Posch, G. Sch\"{u}tz, H. Zhao for highly useful discussions. Fujihira Yuta kindly pointed out inaccuracies in previous versions.
Support by Fund For Math is acknowledged.


\begin{appendix}
\section{Appendix: Coupling constants}\label{app.A}
\setcounter{equation}{0}
\subsection{Equilibrium susceptibilities, currents}\label{A.a}

Averages with respect to $Z^{-1}\exp[-\beta(V(y)+py)]dy$ are denoted in this appendix by $\langle\cdot\rangle$ with the dependence on $p,\beta$ being suppressed. $\langle X_1;X_2\rangle = \langle X_1 X_2\rangle - \langle X_1\rangle\langle X_2\rangle$ is the second cumulant and
\begin{eqnarray}\label{A.0}
&&\hspace{-20pt}\langle X_1;X_2;X_3\rangle = \langle X_1 X_2 X_3 \rangle\nonumber\\
&&\hspace{0pt}- \langle X_1 X_2\rangle\langle X_3\rangle- \langle X_1 X_3\rangle\langle X_2\rangle- \langle X_2 X_3\rangle\langle X_1\rangle+ 2 \langle X_1\rangle\langle X_2\rangle\langle X_3\rangle
\end{eqnarray}
the third cumulant. It holds
\begin{eqnarray}\label{A.1}
&&\hspace{-10pt}
\partial_p \langle X_1\rangle=-\beta \langle X_1;y\rangle\,,\quad \partial_\beta \langle X_1\rangle=- \langle X_1;V+py\rangle\,,
\nonumber\\
&&\hspace{-10pt} \partial_p \langle X_1;X_2\rangle=-\beta \langle X_1;X_2;y\rangle\,,\quad \partial_\beta \langle X_1;X_2\rangle=- \langle X_1;X_2;V+py\rangle\,.
\end{eqnarray}
In equilibrium, $\{r_j,p_j, j\in \mathbb{Z}\}$ are independent  random variables. $p_j$ has a Gaussian den\-sity with mean zero
and variance $\beta^{-1}$ and $r_j$ has the density $Z^{-1}\exp[-\beta(V(y)+py)]$. Hence $S(j,0) = \delta_{0j} C$ with 
\begin{equation}\label{A.2}
C=
\begin{pmatrix} \langle y;y\rangle & 0 & \langle y;V\rangle \\
               0 & \beta^{-1} & 0 \\
                \langle y;V\rangle & 0 & \tfrac{1}{2}\beta^{-2}+\langle V;V\rangle
\end{pmatrix}\,.
\end{equation}
The macroscopic conserved fields are $(\ell,\mathsf{u},\mathfrak{e})$. Their currents are
\begin{equation}\label{A.3}
\big(-\mathsf{u},p(\ell,\mathfrak{e}-\tfrac{1}{2} \mathsf{u}^2), \mathsf{u}p(\ell,\mathfrak{e}-\tfrac{1}{2}\mathsf{u}^2)\big)\,.
\end{equation}
We linearize the Euler equations at $(\ell, \mathsf{u}=0,\mathsf{e})$, $\mathsf{e}$ the internal energy, where
\begin{equation}\label{A.4}
\ell =\langle y\rangle\,,\quad \mathsf{e}= \tfrac{1}{2}\beta^{-1} + \langle V\rangle\,,
\end{equation}
which defines $(p,\beta) \mapsto (\ell (p,\beta),\mathsf{e}(p,\beta))$.  Inverting this map yields
\begin{equation}\label{A.5}
p=p(\ell,\mathsf{e})\,,\quad \beta=\beta(\ell,\mathsf{e})\,.
\end{equation}.


\subsection{Linearization, normal modes}\label{A.b}

The linearized currents are
\begin{equation}\label{A.6}
A=
\begin{pmatrix} 0 & -1 & 0 \\
               \partial_\ell p & 0 & \partial_\mathsf{e}p \\
                0 & p & 0
\end{pmatrix}
\end{equation}
with the property that
\begin{equation}\label{A.6a}
  AC=CA^{\mathrm{T}}\,.
\end{equation}
$A$ has the eigenvalues $c_\sigma=\sigma c$, $c_0=0$, $\sigma=\pm 1$, $c$ the sound speed,
\begin{equation}\label{A.7}
c^2= -\partial_{\ell} p+p \partial_\mathsf{e}p>0\,.
\end{equation}
$A$ has right eigenvectors defined by $A|\psi_\alpha\rangle = c_\alpha |\psi_\alpha\rangle$ and given by
\begin{equation}\label{A.8}
\psi_0=Z^{-1}_0
\begin{pmatrix} \partial_\mathsf{e} p\\
                0\\
                -\partial_\ell p \\
\end{pmatrix}\,,\quad \psi_\sigma=Z^{-1}_\sigma
\begin{pmatrix} -1\\
                \sigma c\\
                p \\
\end{pmatrix}
\end{equation}
and left eigenvectors defined by $\langle\tilde{\psi}_\alpha|A = c_\alpha \langle\tilde{\psi}_\alpha|$ and given by
\begin{equation}\label{A.9}
\tilde{\psi}_0=\tilde{Z}^{-1}_0
\begin{pmatrix} p\\
                0\\
                1 \\
\end{pmatrix}\,,\quad \tilde{\psi}_\sigma=\tilde{Z}^{-1}_\sigma
\begin{pmatrix} \partial_\ell p \\
                \sigma c\\
                \partial_\mathsf{e} p \\
\end{pmatrix}\,,
\end{equation}
which satisfy $\langle \tilde{\psi}_\alpha|\psi_\beta\rangle=0$ for $\alpha\neq\beta$. The linear transformation to normal modes, $\vec{\phi}=R\vec{u}$, is defined through
\begin{equation}\label{A.10}
RAR^{-1}= \mathrm{diag}(-c,0,c)\,,\quad RCR^\mathrm{T}=1\,.
\end{equation}
The first identity can be achieved by setting
\begin{eqnarray}\label{A.12}
&&\hspace{-40pt}R=
\begin{pmatrix} \langle\tilde{\psi}_-|\\
                \langle\tilde{\psi}_0|\\
                \langle\tilde{\psi}_+| \\
\end{pmatrix} =
( \tilde{Z}_{1})^{-1}\begin{pmatrix}
\partial_\ell p&- c&\partial_ \mathsf{e}p\\
  \tilde{\kappa}p&0& \tilde{\kappa}\\
\partial_\ell p& c &\partial_ \mathsf{e}p\\
\end{pmatrix}\,,\\[1ex]
&&\hspace{-40pt}R^{-1}=\big(|\psi_-\rangle |\psi_0\rangle |\psi_+\rangle\big)
= (Z_{1})^{-1}\begin{pmatrix}
-1&\kappa\partial_ \mathsf{e}p&- 1\\ - c&0&c
\\ p&-\kappa\partial_\ell p&p
\end{pmatrix}\,.
\end{eqnarray}
The normalization factors are still free, but up to an overall factor of $-1$ determined by the second identity of~(\ref{A.10}),
\begin{eqnarray}\label{A.11}
&&\hspace{-20pt}Z_0\tilde{Z}_0=c^2,\quad \tilde{Z}^2_0=\tfrac{1}{2}\beta^{-2}+\langle V+py;V+py\rangle = \Gamma c^2\,, \nonumber\\
&&\hspace{-20pt}Z_\sigma\tilde{Z}_\sigma=2c^2,\quad  \tilde{Z}^2_\sigma=2 \beta^{-1}c^2\,,\quad
\kappa = Z_1/Z_0 \,, \quad\tilde{\kappa} =  \tilde{Z}_1/ \tilde{Z}_0\,.
\end{eqnarray}

The Euler part of the equations of motion reads
\begin{equation}
\partial_t u_\alpha + \partial _x\big((A\vec{u})_\alpha + \tfrac{1}{2} \langle\vec{u}, H^{\alpha}\vec{u}\rangle\big)= 0\,.
\end{equation}
Using $\vec{\phi} = R \vec{u}$, one arrives at
\begin{equation}
\partial_t \phi_\alpha + \partial _x\big(c_\alpha \phi_\alpha + \tfrac{1}{2}\sum_{\alpha' =1}^3 R_{\alpha\alpha'} \langle R^{-1}\vec{\phi}, H^{\alpha'}
R^{-1}\vec{\phi}\rangle\big)= 0\,,
\end{equation}
which implies
\begin{equation}
\partial_t \phi_\alpha + \partial _x\big( c_\alpha\phi_\alpha +  \langle\vec{\phi}, G^{\alpha}\vec{\phi}\rangle\big)= 0
\end{equation}
with the normal mode coupling constants
\begin{equation}\label{A.14}
G^\alpha_{\beta\gamma}=\tfrac{1}{2}\sum_{\alpha'=1}^3R_{\alpha\alpha'} 
\langle\psi_\beta, H^{\alpha'} \psi_\gamma\rangle\,.
\end{equation}


\subsection{Hessians and $G$ couplings}\label{A.c}

The three Hessians are
\begin{equation}\label{A.13}
H^\ell =0\,,\quad H^\mathsf{u}=
\begin{pmatrix} \partial^2_\ell p & 0 & \partial_\ell \partial_\mathsf{e}p \\
               0 & -\partial_\mathsf{e}p & 0 \\
                \partial_\ell \partial_\mathsf{e}p & 0 & \partial^2_\mathsf{e}p
\end{pmatrix}\,,\quad H^\mathsf{e}=
\begin{pmatrix} 0 & \partial_\ell p & 0 \\
               \partial_\ell p & 0 & \partial_\mathsf{e}p \\
                0 & \partial_\mathsf{e}p & 0
\end{pmatrix}\,.
\end{equation}
The $H^\mathsf{u}$ matrix elements are given by
\begin{eqnarray}\label{A.15}
&&\hspace{-20pt}\langle\psi_0,H^\mathsf{u} \psi_0\rangle=\frac{1}{Z^2_0} \big(\partial^2_\ell p\partial_\mathsf{e}p \partial_\mathsf{e}p-2\partial_\ell\partial_\mathsf{e}p \partial_\ell p \partial_\mathsf{e}p +\partial^2_\mathsf{e}p \partial_\ell p\partial_\ell p\big)\,,
\\
&&\hspace{-20pt}\langle\psi_0,H^\mathsf{u} \psi_\sigma\rangle=\frac{1}{Z_0 Z_\sigma} \big(-\partial^2_\ell p\partial_\mathsf{e}p-p \partial^2_\mathsf{e}p\partial_\ell p +p \partial_\ell \partial_\mathsf{e}p \partial_\mathsf{e}p +\partial_\ell \partial_\mathsf{e} p 
\partial_\ell p\big)\,,\\
&&\hspace{-20pt}\langle\psi_{\sigma},H^\mathsf{u} \psi_{\sigma'}\rangle=\frac{1}{Z_\sigma Z_{\sigma'}} \big(\partial^2_\ell p- 2p\partial_\ell\partial_\mathsf{e}p +p^2 \partial^2_\mathsf{e}p-\sigma\sigma' c^2  \partial_\mathsf{e}p \big)
\end{eqnarray}
and the $H^\mathsf{e}$ matrix elements are
\begin{equation}\label{A.16}
\langle\psi_0,H^\mathsf{e} \psi_0\rangle = 0\,,\quad
\langle\psi_0,H^\mathsf{e}\psi_\sigma\rangle = 0\,,\quad
\langle\psi_{\sigma},H^\mathsf{e} \psi_{\sigma'}\rangle = c(2\beta)^{-1}(\sigma + \sigma')\,.
\end{equation}
Denoting the standard basis vectors by $\mathrm{e}_1,\mathrm{e}_2,\mathrm{e}_3$, one has
\begin{equation}\label{A.17}
R\mathrm{e}_2 = (\tfrac{1}{2}\beta)^{1/2}(- \mathrm{e}_1 + \mathrm{e}_3)\,,\quad
R\mathrm{e}_3 = c^{-1}(\tfrac{1}{2}\beta)^{1/2}\partial_\mathsf{e}p (\mathrm{e}_1 + \mathrm{e}_3)
+ (\tilde{Z}_0)^{-1} \mathrm{e}_2
   \end{equation}
and arrives at the coupling constants
\begin{eqnarray}\label{A.18}
&&\hspace{-10pt}G^\sigma_{\beta\gamma}=\tfrac{1}{2}\sigma (\tfrac{1}{2}\beta)^{1/2} \langle \psi_\beta,H^\mathsf{u}\psi_\gamma\rangle+
\tfrac{1}{2}\partial_\mathsf{e}p (\tfrac{1}{2}\beta)^{1/2}c^{-1} \langle \psi_\beta,H^\mathsf{e}\psi_\gamma\rangle\,,\\\label{A.19}
&&\hspace{-10pt}G^0_{\beta\gamma}=\frac{1}{2\tilde{Z}_0} \langle \psi_\beta,H^\mathsf{e}\psi_\gamma\rangle\,.
\end{eqnarray}
They have the symmetries
\begin{eqnarray}\label{A.20}
&&\hspace{-20pt}G^\alpha_{\beta\gamma} = G^\alpha_{\gamma\beta}\,, \quad G^\sigma_{\alpha\beta} 
= - G^{-\sigma}_{-\alpha-\beta}\,,\quad
 G^\sigma_{-10}  = G^\sigma_{01}  \,,  \nonumber\\
 &&\hspace{-20pt}G^0_{\sigma\sigma}  = - G^0_{-\sigma-\sigma}\,,\quad  
G^0_{\alpha\beta} = 0\,\, \mathrm{otherwise} \,.
\end{eqnarray}


\subsection{Transformation to canonical variables}\label{A.d}

We transform the couplings $G$ from microcanonical to canonical variables. In canonical variables, all $G$ coefficients are given by up to third order cumulants in $y,V,V+py$, which can be completely expressed in terms of one-dimensional integrals. A Mathematica program computes all $G$ coefficients for specified $V,p,\beta$ and by the same procedure also the $R,R^{-1}$
matrices.

Differentiating the identities $p(\ell(p,\beta),\mathsf{e}(p,\beta))= p$ and $\beta(\ell(p,\beta),\mathsf{e}(p,\beta))= \beta$
with respect to $p$ and $\beta$ yields
\begin{equation}\label{A.20a}
\begin{pmatrix}
  1 \\
  0 \\
\end{pmatrix}=
\begin{pmatrix}
  \partial_p\ell & \partial_p\mathsf{e} \\
  \partial_\beta\ell & \partial_\beta\mathsf{e} \\
\end{pmatrix}
\begin{pmatrix}
  \partial_\ell p \\
  \partial_\mathsf{e} p \\
\end{pmatrix}\,,
\end{equation}
\begin{equation}\label{A.21}
\begin{pmatrix}
  0 \\
  1\\
\end{pmatrix}=
\begin{pmatrix}
  \partial_p\ell & \partial_p\mathsf{e} \\
  \partial_\beta\ell & \partial_\beta\mathsf{e} \\
\end{pmatrix}
\begin{pmatrix}
  \partial_\ell \beta \\
  \partial_\mathsf{e} \beta \\
\end{pmatrix}\,.
\end{equation}
Hence
\begin{eqnarray}\label{A.22}
&&\hspace{-3pt}
\partial_\ell p = \Gamma^{-1}\partial_\beta \mathsf{e} =-\Gamma^{-1} \big( \tfrac{1}{2}\beta^{-2} +\langle V;V+py\rangle\big)\,,
\nonumber\\[1ex]
&&\hspace{-3pt} \partial_\mathsf{e} p = -\Gamma^{-1}\partial_\beta \ell =\Gamma^{-1}\langle y;V+py\rangle\,,\nonumber\\[1ex]
&&\hspace{7pt} \Gamma= \beta \big(\langle y;y\rangle\langle V;V\rangle - \langle y;V\rangle^2\big) +\tfrac{1}{2}\beta^{-1}\langle y;y\rangle\,,
\end{eqnarray}
and for the velocity of sound
\begin{equation}\label{A.23}
c^2= \frac{1}{\Gamma}\big( \tfrac{1}{2}\beta^{-2}+\langle V+py;V+py\rangle\big)\,.
\end{equation}

We collect first and second derivatives of $\ell,\mathsf{e}$ and first derivatives of $\Gamma$,
\begin{eqnarray}
&&\hspace{-45pt} \partial_p \ell = - \beta \langle y;y\rangle\,,\quad  \partial_\beta \ell =  - \langle y;V + py\rangle\,,\\[1ex]
&&\hspace{-45pt}\partial_p \mathsf{e} = -\beta \langle y;V\rangle\,,\quad \partial_\beta\mathsf{e}= - \tfrac{1}{2} \beta^{-2}- 
\langle V;V +py \rangle\,,\\[1ex]
&&\hspace{-45pt}\partial_p\partial_\beta \ell = -  \langle y;y\rangle +\beta \langle y;y;V+ py\rangle\,,\quad \partial_\beta ^2
\ell = \langle y;V+py;V +py\rangle\,,\\[1ex]
&&\hspace{-45pt}\partial_p\partial_\beta  \mathsf{e} = - \langle y;V\rangle + \beta \langle y;V; V +py\rangle\,, \quad \partial_\beta^2\mathsf{e}= \beta^{-3} + \langle V;V+py;V+py\rangle\,,\\[1ex]
&&\hspace{-45pt}\partial_p \Gamma = \beta^2\big( -\langle y;y;y\rangle\langle V;V\rangle  - \langle y;y\rangle\langle y;V;V\rangle
+ 2 \langle y;V\rangle\langle y;y;V\rangle\big) -\tfrac{1}{2}\langle y;y;y\rangle\,,\\[1ex]
&&\hspace{-45pt}\partial_\beta \Gamma = \langle y;y\rangle \langle V;V\rangle - \langle y;V\rangle^2 - \tfrac{1}{2}\beta^{-2}\langle y;y\rangle +\beta\big( - \langle y;y;V+py\rangle \langle V;V\rangle\nonumber\\
&&\hspace{-22pt}  -  \langle y;y\rangle \langle V;V;V+py\rangle +2 \langle y;V\rangle \langle y;V;V+py\rangle\big)
- \tfrac{1}{2} \beta^{-1} \langle y;y;V+py\rangle  \,.
\end{eqnarray}

To complete our task we still need the second derivatives $\partial_\ell\partial_\ell p$, $\partial_\ell\partial_\mathsf{e} p$, $\partial_\mathrm{e}\partial_\mathsf{e} p$, where we start from
\begin{equation}\label{A.29}
\begin{pmatrix}
  \partial_p(\Gamma^{-1} \partial_\beta \mathsf{e}) \\
  \partial_\beta(\Gamma^{-1}  \partial_\beta \mathsf{e} )\\
\end{pmatrix}=
\begin{pmatrix}
  \partial_p\ell & \partial_p\mathsf{e} \\
  \partial_\beta\ell & \partial_\beta\mathsf{e} \\
\end{pmatrix}
\begin{pmatrix}
  \partial_\ell  \partial_\ell p \\
  \partial_\mathsf{e}  \partial_\ell p \\
\end{pmatrix}\,,
\end{equation}

\begin{equation}\label{A.30}
-\begin{pmatrix}
  \partial_p(\Gamma^{-1} \partial_\beta \ell )\\
  \partial_\beta( \Gamma^{-1} \partial_\beta \ell )\\
\end{pmatrix}=
\begin{pmatrix}
  \partial_p\ell & \partial_p\mathsf{e} \\
  \partial_\beta\ell & \partial_\beta\mathsf{e} \\
\end{pmatrix}
\begin{pmatrix}
  \partial_\ell  \partial_\mathsf{e} p \\
  \partial_\mathsf{e}  \partial_\mathsf{e} p \\
\end{pmatrix}\,.
\end{equation}
Inverting~(\ref{A.29}),~(\ref{A.30}) we arrive at
\begin{eqnarray}
&&\hspace{-17pt}\partial^2_\ell p= - \Gamma^{-2}(\partial_p\mathsf{e}\partial_\beta^2\mathsf{e} -\partial_\beta\mathsf{e}
\partial_p\partial_\beta\mathsf{e} )+ \Gamma^{-3}(\partial_p\mathsf{e}\partial_\beta \mathsf{e}\partial_\beta\Gamma
-\partial_\beta\mathsf{e}\partial_\beta \mathsf{e}\partial_p\Gamma)\,,\\[1ex]
&&\hspace{-17pt}\partial_\ell\partial_\mathsf{e} p = -\Gamma^{-2}(- \partial_p\mathsf{e} 
\partial_\beta^2 \mathsf{\ell}+
\partial_\beta\mathsf{e} \partial_p\partial_\beta\ell) + \Gamma^{-3}(-\partial_p\mathsf{e} \partial_\beta\ell \partial_\beta \Gamma + \partial_\beta \mathsf{e} \partial_\beta\ell\partial_p\Gamma)\,,\\[1ex]
&&\hspace{-17pt}\partial_\mathsf{e}^2p = \Gamma^{-2}(- \partial_p\ell\partial_\beta^2\ell + \partial_\beta \ell \partial_p\partial_\beta\ell) - \Gamma^{-3}(-\partial_\beta \ell \partial_p \ell \partial_\beta\Gamma + \partial_\beta\ell \partial_\beta \ell \partial_p\Gamma)\,.
\end{eqnarray}
By successive substitutions, the coupling constants in~(\ref{A.18}),~(\ref{A.19}) are expressed in terms of cumulants 
in $y, V, V+py$ at most of order three.

\subsection{Even potential, zero pressure}\label{A.e}

For an even potential and $p=0$ by symmetry $\langle y;V\rangle=0$, $\langle y;V;V\rangle=0$, $\langle y;y;y\rangle=0$. This simplifies the expressions for the couplings and $c^2$. It holds
\begin{equation}\label{A.32}
c^2 = (\beta\langle y;y\rangle)^{-1}\,.
\end{equation}
The only non-zero couplings are
\begin{eqnarray}\label{A.31}
&&\hspace{-13pt}G^\sigma_{0 \sigma'} = G^\sigma_{\sigma' 0}
=\tfrac{1}{4}\sigma c^3\big((\beta c^2)^{-1} - \beta\langle y;y;V \rangle\big) \big((2\beta^2)^{-1}+ \langle V;V\rangle\big)^{-1/2}\,,\quad \sigma,\sigma'=\pm 1\,,\nonumber\\
&&\hspace{-13pt} G^0_{\sigma \sigma} = \tfrac{1}{2}\sigma c \beta^{-1} \big((2\beta^2)^{-1}+ \langle V;V\rangle\big)^{-1/2}\,.
\end{eqnarray}

\subsection{Second sum rule}\label{A.f}
As claimed in Eq.~(\ref{23}), the linearized Euler currents, $A$, and the susceptibility matrix, $C$, satisfy
\begin{equation}\label{a1}
AC = CA^\mathrm{T}.
\end{equation}
This relation is well known for classical fluids, see \textit{e.g.}~\cite{Spo91}. In fact, (\ref{a1}) is very general and relies only on
space-time stationarity. Let us denote, independently of any particular model,  the conserved fields by  $\eta_\alpha(j,t)$, $j\in\mathbb{Z}$, $t \in\mathbb{R}$, $\alpha = 1,\ldots,n$, which are assumed to be
space-time stationary with zero mean. The corresponding currents are denoted by $\mathcal{J}_\alpha(j,t)$. By stationarity
\begin{equation}\label{App1}
S_{\alpha\beta}(j,t) = \langle  \eta_\alpha(j,t) \eta_\beta(0,0)\rangle = S_{\beta\alpha}(-j,-t)\,.
\end{equation}

Using the conservation law,
\begin{equation}
\begin{aligned}
 \frac{d}{dt}\sum_{j \in \mathbb{Z}} jS_{\alpha\beta}(j,t) &=  \sum_{j \in \mathbb{Z}} j
\langle   (\mathcal{J}_\alpha(j-1,t) -   \mathcal{J}_\alpha(j,t))\eta_\beta(0,0)\rangle \\
&= \sum_{j \in \mathbb{Z}} \langle   \mathcal{J}_\alpha(j,t)\eta_\beta(0,0)\rangle =   \sum_{j \in \mathbb{Z}} \langle   \mathcal{J}_\alpha(0,0)\eta_\beta(-j,-t)\rangle \\
&= \sum_{j \in \mathbb{Z}} \langle   \mathcal{J}_\alpha(0,0)\eta_\beta(j,0)\rangle\,.
\end{aligned}
\end{equation}
As standard for mechanical systems in thermal equilibrium, but also valid for stochastic lattice gases with conservation laws
\cite{Ku84,GSch12}, the infinite volume average can be obtained from a system on a ring by introducing
a chemical potential, $\mu_\alpha$, for the density $\rho_\alpha$.  Hence
\begin{eqnarray}\label{App2}
&&\hspace{0pt}\sum_{j \in \mathbb{Z}} \langle   \mathcal{J}_\alpha(0,0)\eta_\beta(j,0)\rangle = 
\frac{\partial}{\partial \mu_{\beta}} \langle   \mathcal{J}_\alpha(0,0)\rangle_{\vec{\mu}}\\
&&\hspace{0pt}= \sum_{\gamma =1}^{n} \frac{\partial}{\partial \rho_{\gamma}} \langle   \mathcal{J}_\alpha(0,0)\rangle_{\vec{\rho}} \,\frac{\partial \rho_{\gamma}}{\partial \mu_{\beta}} = (AC)_{\alpha\beta}
\end{eqnarray}
and
\begin{equation}
\sum_{j \in \mathbb{Z}} jS_{\alpha\beta}(j,t) =  (AC)_{\alpha\beta}t +\sum_{j \in \mathbb{Z}} jS_{\alpha\beta}(j,0)\,.
\end{equation}
Summing in Eq.~(\ref{App1}) over $j$ yields
\begin{equation}
(AC)_{\alpha\beta}t = (AC)_{\beta\alpha}t\,,
\end{equation}
which is the desired identity.

For many-component lattice gases Eq.~(\ref{23}) was noted by T\'{o}th and Valk\'{o}~\cite{TV03} in a special case and proved in generality  by Grisi and Sch\"{u}tz~\cite{GSch12}. 

\section{The hard-point gas with alternating masses}\label{A.h}
\setcounter{equation}{0}

As the name suggests, the hard-point gas consists of point particles which collide elastically. This amounts to merely exchanging the labels and thus ideal gas dynamics. To introduce more chaotic elements one modifies the model to 
have alternating masses, say $m_0$ and $m_1$. Numerically the choice $m_1/m_0 = 3$ seems to have good time mixing. The simulation runs from collision to collision which is much faster than solving differential equations.
Regarded as a chain, the hard-point gas has maximal simplicity and appears to be a favorable candidate the check our predictions. At the same time it serves as a nice illustration of the method.

As a novel feature, the unit cell consists of two particles. Let us first consider the case of hard-points with a general potential $V$
 and let us reintroduce the mass $m$ of a particle. Then the hydrodynamic currents from (\ref{16}) are modified as 
\begin{equation}\label{A.40}
-\tfrac{1}{m}\mathsf{u}, p, \tfrac{1}{m}\mathsf{u}p\,, \quad p = p(\ell,\mathfrak{e} - \tfrac{1}{2m}\mathsf{u}^2)\,,
\end{equation}
$\mathsf{u}$ the momentum density. Alternating masses modify the currents. We claim that in (\ref{A.40})
one merely has to substitute  for $m$ the average mass
\begin{equation}\label{A.41}
\bar{m} = \tfrac{1}{2}(m_0 +m _1)\,.
\end{equation}
To verify the claim we use the relation $AC = CA$, which holds in generality. The static correlator $C$ is modified to 
\begin{equation}\label{A.42}
C=
\begin{pmatrix} \langle y;y\rangle & 0 & \langle y;V\rangle \\
               0 & \beta^{-1}\bar{m} & 0 \\
                \langle y;V\rangle & 0 & \tfrac{1}{2}\beta^{-2}+\langle V;V\rangle
\end{pmatrix}\,,
\end{equation}
since
\begin{equation}\label{A.43}
\tfrac{1}{2}\langle(p_0 + p_1)^2\rangle = \beta^{-1}\bar{m}\,.
\end{equation}
Hence the linearization $A$ is modified to
\begin{equation}\label{A.44}
A=
\begin{pmatrix} 0 & -\tfrac{1}{\bar{m}} & 0 \\
               \partial_\ell p & 0 & \partial_\mathsf{e}p \\
                0 & \tfrac{1}{\bar{m}}p & 0
\end{pmatrix}
\end{equation}
as claimed. 

The hard-point gas has ideal gas thermodynamics, which means
\begin{equation}\label{A.45}
p = \frac{\mathsf{2e}}{\ell}\,,\, \beta = \frac{1}{2\mathsf{e}}\,.
\end{equation}
Thus $\partial_\ell p= -\beta p^2$, $\partial_\mathsf{e}p= 2\beta p$ and the sound speed reads
\begin{equation}\label{A.46}
c_{\bar{m}} = (3\beta/\bar{m})^{1/2}p\,.
\end{equation}
The transformation matrix is obtained to 
\begin{equation}\label{A.47}
R = 
\frac{1}{\sqrt{6}}\begin{pmatrix}
-\beta p&-  \sqrt{3\beta/\bar{m}}&2\beta\\
2\beta p&0& 2\beta \\
-\beta p& \sqrt{3\beta/\bar{m}}&2\beta\\
\end{pmatrix}\,,
\end{equation}
\begin{equation}\label{A.48}
R^{-1}= \frac{1}{\sqrt{6}\beta p}
\begin{pmatrix}
-1&2&-1\\
-\sqrt{3\beta\bar{m}} \,p&0& \sqrt{3\beta \bar{m}} \,p\\
p&p&p\\
\end{pmatrix}\,.
\end{equation}

The correlations of the physical fields are then given through $S= R^{-1}S^\sharp R^{-1\mathrm{T}}$, where $S^\sharp$
is assumed to be approximately diagonal,
\begin{equation}\label{A.49}
S^\sharp_{\alpha\beta} = \delta_{\alpha\beta}f_\alpha\,.
\end{equation}
 Using (\ref{A.48})
one obtains
\begin{eqnarray}\label{A.50}
&&\hspace{0pt}\ell -\ell \;\;\mathrm{correlations}:\quad\frac{1}{6\beta^2 p^2}(f_{-1} + 4 f_0 +f_1)\,,\nonumber\\
&&\hspace{0pt}\mathsf{u}-\mathsf{u} \;\;\mathrm{correlations}:\quad \frac{\bar{m}}{2\beta}(f_{-1} + f_1)\,,\nonumber\\
&&\hspace{0pt}\mathsf{e}-\mathsf{e} \;\;\mathrm{correlations}:\quad \frac{1}{6\beta ^2}(f_{-1} +  f_0 +f_1)\,.\nonumber\\
\end{eqnarray}

Next we compute the $G$ matrices. Firstly, by direct differentiation of $p$,
\begin{equation}\label{A.51}
H^\ell =0\,,\quad H^\mathsf{u}= \frac{2}{\ell^3}
\begin{pmatrix} 2 \mathsf{e} & 0 & -\ell \\
               0 & -\bar{m}^{-1}\ell^2 & 0 \\
                -\ell & 0 & 0
\end{pmatrix}\,,\quad H^\mathsf{e}= \frac{2}{\bar{m} \ell^2}
\begin{pmatrix} 0 & - \mathsf{e}& 0 \\
              - \mathsf{e} & 0 & \ell \\
                0 & \ell & 0
\end{pmatrix}
\end{equation}
and transformed as
\begin{equation}\label{A.52}
(R^{-1})^{\mathrm{T}}H^\mathsf{u} R^{-1}= p D_0\,, \quad
  (R^{-1})^{\mathrm{T}}H^\mathsf{e} R^{-1} = \frac{c_{\bar{m}}}{\beta} D_1
\end{equation}
with the matrices
\begin{equation}\label{A.51a} D_0= 
\begin{pmatrix} 0 & -1 & 2 \\
               -1 & 0 & -1 \\
                2 & -1 & 0
\end{pmatrix}\,,\quad D_1= 
\begin{pmatrix} -1 & 0& 0 \\
               0 & 0 & 0 \\
                0 & 0 & 1
\end{pmatrix}\,.
\end{equation}
We conclude that 
\begin{equation}\label{A.52a}
G^{\pm1} = \frac{c_{\bar{m}}}{2 \sqrt{6}}(\pm D_0  + 2D_1)\,, \quad G^0 = \frac{c_{\bar{m}}}{ \sqrt{6}} D_1
\end{equation}
and more explicitly
 \begin{equation}\label{A.53} 
 G^{\pm1} = \pm\frac{c_{\bar{m}}}{ 2\sqrt{6}}
\begin{pmatrix} \mp 2 & -1 & 2 \\
               -1 & 0 & -1 \\
                2 & -1 & \pm 2
\end{pmatrix}\,,\quad G^0= \frac{c_{\bar{m}}}{ \sqrt{6}}
\begin{pmatrix} -1 & 0& 0 \\
               0 & 0 & 0 \\
                0 & 0 & 1
\end{pmatrix}\,.
\end{equation} 
 
As a special feature of the hard-point gas, all non-universal parameters are accounted for by the sound speed 
$c_{\bar{m}}$. Most importantly, it should 
be noted that $G^1_{00} = 0$, which implies that to leading order there is no coupling of the heat mode to the sound
mode. This is an indication that for the sound mode peaks the finite time corrections are less severe than in 
asymmetric FPU chains.


\section{Appendix: Discretized multi-component KPZ, \\mode-coupling}\label{app.B}
\setcounter{equation}{0}
\subsection{Discretized KPZ}\label{B.a}

The Langevin equation~(\ref{34}) is somewhat formal. To have a well-defined evolution, we lattice discretize space by a lattice of $N$ sites. The field $\phi(x,t)$ then becomes $\phi_j(t)$ with components $\phi_{j,\alpha}(t)$, $j=1,\ldots,N$, $\alpha=1,\ldots,n$. The spatial  finite difference operator is denoted by $\tau$, $\tau f_j=f_j-f_{j-1}$, with transpose $\tau^\mathrm{T} f_j=f_j-f_{j+1}$. Then the discretized KPZ equation reads
\begin{equation}\label{B.1}
\partial_t \phi_{j,\alpha}+\tau\big(c_\alpha \phi_{j,\alpha} +\mathcal{N}_{j,\alpha} + \tau^\mathrm{T} D\phi_{j,\alpha} + \sqrt{2D}\xi_{j,\alpha}\big)=0
\end{equation}
with $\phi_{0}=\phi_N$, $\phi_{N+1}=\phi_1$, $\xi_0=\xi_N$, where $\xi_{j,\alpha}$ are independent Gaussian white noises with covariance
\begin{equation}\label{B.2}
\langle \xi_{j,\alpha}(t) \xi_{j',\alpha'} (t')\rangle =\delta_{jj'} \delta_{\alpha\alpha'} \delta(t-t')\,.
\end{equation}
The diffusion matrix $D$ acts on components, while the difference $\tau$ acts on the lattice site variable $j$.

$\mathcal{N}_{j,\alpha}$ is quadratic in $\phi$. But let us first consider the case $\mathcal{N}_{j,\alpha} =0$. Then, since
according to ~(\ref{B.1}) the drift is linear in $\phi$,  $\phi_{j,\alpha}(t)$ a Gaussian process. The noise strength has been chosen such that one invariant measure is the Gaussian
\begin{equation}\label{B.3}
\prod^N_{j=1} \prod^n_{\alpha=1} \big(\exp[-\tfrac{1}{2}\phi^2_{j,\alpha}] (2\pi)^{-1/2} d \phi_{j,\alpha}\big)= \rho_\mathrm{G} (\phi) \prod^N_{j=1} \prod^n_{\alpha=1} d \phi_{j,\alpha}\,.
\end{equation}
The set of all extremal invariant measures are obtained by conditioning~(\ref{B.3}) on the hyperplanes
\begin{equation}\label{B.4}
\sum^N_{j=1} \phi_{j,\alpha}=N\rho_\alpha\,,
\end{equation}
which for large $N$ would become independent Gaussians with mean $\rho_\alpha$. In the following we fix $\rho_\alpha=0$ and denote the average with respect to $\rho_\mathrm{G}$ by $\langle\cdot\rangle_{\mathrm{eq}}$.

The generator corresponding to~(\ref{B.1}) with $\mathcal{N}_{j,\alpha}=0$ is given by
\begin{equation}\label{B.5}
L_0=\sum^N_{j=1} \Big(-\sum^n_{\alpha=1} \tau\big(c_\alpha \phi_{j,\alpha}+\tau^\mathrm{T} D\phi_{j,\alpha}\big) \partial_{\phi_{j,\alpha}}+\sum^n_{\alpha,\beta=1} 2D_{\alpha\beta} (\tau \partial_{\phi_{j,\alpha}} )(\tau\partial_{\phi_{j,\beta}})\Big)\,.
\end{equation}
The invariance of $\rho_\mathrm{G}(\phi)$ can be checked through
\begin{equation}\label{B.7}
L^\ast_0 \rho_\mathrm{G}(\phi)=0\,,
\end{equation}
where $^\ast$ is the adjoint with respect to the flat volume measure. Furthermore linear functions evolve to linear functions according to
\begin{equation}\label{B.6}
\mathrm{e}^{L_0 t}\phi_{j,\alpha}= \sum^N_{j'=1} \sum^n_{\alpha'=1} (\mathrm{e}^{tB})_{j\alpha,j'\alpha'} \phi_{j',\alpha'}\,,
\end{equation}
where the matrix $B=-\tau\otimes \mathrm{diag} (c_1,\ldots, c_n) -\tau\tau ^\mathrm{T} \otimes D$.

We now add the nonlinearity $\mathcal{N}_{j,\alpha}$. In general, this will modify the invariant measure and we have little control how. Therefore we propose to choose $\mathcal{N}_{j,\alpha}$ such that $\rho_\mathrm{G}$ is left invariant under 
the deterministic flow generated the evolution equation $\tfrac{d}{dt}\phi= -\tau\mathcal{N}$, i.e. under the generator
\begin{equation}\label{B.9}
L_1= - \sum^N_{j=1} \sum^n_{\alpha=1} \tau \mathcal{N}_{j,\alpha} \partial_{\phi_{j,\alpha}}\,.
\end{equation}
The invariance of $\rho_\mathrm{G}$ under $L_1$ is equivalent to the condition
\begin{equation}\label{B.10}
\sum^N_{j=1} \sum^n_{\alpha=1} \phi_{j,\alpha} \tau \mathcal{N}_{j,\alpha}=0\,.
\end{equation}
If $\mathcal{N}_{j,\alpha}$ depends only on the field at sites $j$ and $j+1$, then the most general solution to~(\ref{B.10}) reads
\begin{equation}\label{B.8}
\mathcal{N}_{j,\alpha}=\tfrac{1}{3}\sum^n_{\beta,\gamma=1} G^\alpha_{\beta\gamma}\big(\phi_{j,\beta}\phi_{j,\gamma}+\phi_{j,\beta}\phi_{j+1,\gamma}+\phi_{j+1,\beta}\phi_{j+1,\gamma}\big)
\end{equation}
under the constraint that
\begin{equation}\label{B.9a}
G^\alpha_{\beta\gamma}=G^\alpha_{\gamma\beta}=G^\beta_{\alpha\gamma}
\end{equation}
for all $\alpha,\beta,\gamma=1,\ldots,n$. Denoting the generator of the Langevin equation~(\ref{B.1}) by
\begin{equation}\label{B.10a}
  L=L_0+L_1\,,
\end{equation}
one concludes $L^\ast \rho_\mathrm{G}=0$, \textit{i.e.} the invariance of $\rho_\mathrm{G}$.

In the continuum limit the condition~(\ref{B.10}) reads
\begin{equation}\label{B.11}
\sum^n_{\alpha,\beta,\gamma=1} G^\alpha_{\beta\gamma} \int dx\phi_\alpha(x) \partial_x \big(\phi_\beta(x) \phi_\gamma(x)\big)=0\,,
\end{equation}
where $G^\alpha_{\beta\gamma}=G^\alpha_{\gamma\beta}$. By partial integration
\begin{equation}\label{B.13}
2 \sum^n_{\alpha,\beta,\gamma=1} G^\alpha_{\beta\gamma} \int dx \phi_\alpha(x) \phi_\beta(x) \partial_x \phi_\gamma(x)  
 =- \sum^n_{\alpha,\beta,\gamma} G^\alpha_{\beta\gamma} \int dx \phi_\beta(x) \phi_\gamma(x) \partial_x \phi_\alpha(x)
\end{equation}
and~(\ref{B.11}) is satisfied only if $G^\gamma_{\beta\alpha}=G^\alpha_{\beta\gamma}$, which is the condition 
(\ref{B.9a}) obtained already in the discrete setting.

We henceforth assume~(\ref{B.9a}) although it will not hold for anharmonic chains, in general. The leading coefficients $G^\alpha_{\alpha\alpha}$'s are not constrained and one can still freely choose the sub-leading $G^\alpha_{\beta\beta}$. The symmetry thus restricts the coefficients corresponding to sub-sub-leading  terms. Appealing to universality we expect that the true invariant measure for general $G$ will have short range correlations and nonlinear fluctuating hydrodynamics remains 
a valid approximation to the microscopic dynamics.

\subsection{Mode-coupling}\label{B.b}

We consider the stationary $\phi_{j,\alpha}(t)$ process, average denoted by $\langle\cdot\rangle$, governed by~(\ref{B.1})
with $\rho_\mathrm{G}$ as $t=0$ measure.  For $t\geq 0$ the stationary covariance reads
\begin{equation}\label{B.14}
S_{\alpha\beta}(j,t)=\langle \phi_{j,\alpha}(t)\phi_{0,\beta}(0)\rangle =\langle \phi_{0,\beta}\mathrm{e}^{Lt}\phi_{j,\alpha}\rangle_{\mathrm{eq}}\,,
\end{equation}
where for easier reading we leave out the superscript $^{\sharp \phi}$. By construction
\begin{equation}\label{B.15}
S_{\alpha\beta}(j,0)=\delta_{\alpha\beta} \delta_{j0}\,.
\end{equation}
The time derivative reads
\begin{equation}\label{B.16}
\frac{d}{dt} S_{\alpha\beta}(j,t)=\langle \phi_{0,\beta}(\mathrm{e}^{Lt}L_0 \phi_{j,\alpha})\rangle_{\mathrm{eq}}+ \langle \phi_{0,\beta}(\mathrm{e}^{Lt}L_1 \phi_{j,\alpha})\rangle_{\mathrm{eq}}\,.
\end{equation}
We insert
\begin{equation}\label{B.17}
\mathrm{e}^{Lt}=\mathrm{e}^{L_0 t}+\int^t_0 ds\, \mathrm{e}^{L_0(t-s)} L_1 \mathrm{e}^{Ls}
\end{equation}
in the second summand of~(\ref{B.16}). The term containing $\mathrm{e}^{L_0 t}$ does not show, since it is cubic in the time zero fields and the average $\langle\cdot\rangle_\mathrm{eq}$ vanishes. Therefore one arrives at 
\begin{equation}\label{B.18}
\frac{d}{dt} S_{\alpha\beta}(j,t)= \sum_{j'\in\mathbb{Z}} \sum^n_{\alpha'=1} \big(B_{\alpha j,\alpha' j'} S_{\alpha'\beta}(j',t) + \int^t_0 ds \langle \phi_{0,\beta} \mathrm{e}^{L_0(t-s)}L_1(\mathrm{e}^{Ls} L_1 \phi_{j,\alpha}) \rangle_\mathrm{eq}\big)\,.
\end{equation}
For the adjoint of $\mathrm{e}^{L_0(t-s)}$ we use~(\ref{B.6}) and for the adjoint of $L_1$ we use
\begin{equation}\label{B.18a}
\langle \phi_{j,\alpha}  L_1  F(\phi)\rangle_\mathrm{eq} =  - \langle (L_1 \phi_{j,\alpha})   F(\phi)\rangle_\mathrm{eq} \,.
\end{equation}
Furthermore
\begin{equation}\label{B.19}
L_1 \phi_{j,\alpha} = - \tau \mathcal{N}_{j,\alpha}\,.
\end{equation}
Inserting in~(\ref{B.18}) one arrives at the identity
\begin{eqnarray}\label{B.18b}
 &&\hspace{-60pt}\frac{d}{dt} S_{\alpha\beta}(j,t)= \sum_{j'\in\mathbb{Z}} \sum^n_{\alpha'=1} \big(B_{\alpha j,\alpha' j'} S_{\alpha'\beta}(j',t)\nonumber\\
&&\hspace{10pt}  - \int^t_0 ds (\mathrm{e}^{B^\mathrm{T}(t-s)})_{0\beta,j'\alpha'} \langle\tau \mathcal{N}_{j',\alpha'}(\mathrm{e}^{Ls} \tau \mathcal{N}_{j,\alpha}) \rangle_\mathrm{eq}\big)\,.
\end{eqnarray}

To obtain a closed equation for $S$ we note that the average $\langle \tau \mathcal{N}_{j,\alpha}(s)\tau 
\mathcal{N}_{j',\alpha'}(0)\rangle$ appearing in~(\ref{B.18b}) is
a four-point correlation, for which we invoke a Gaussian factorization as
\begin{equation}
\langle\phi(s)\phi(s)\phi(0)\phi(0)\rangle\cong \langle\phi(s)\phi(s)\rangle\langle\phi(0)\phi(0)\rangle+2\langle\phi(s)\phi(0)\rangle\langle\phi(s)\phi(0)\rangle\,.
\end{equation}
 The first summand vanishes because of the difference operator $\tau$. Secondly we replace the bare propagator $\mathrm{e}^{B(t-s)}$ by the interacting propagator $S(t-s)$. Finally we take a limit of zero lattice spacing. This step could be avoided, and is done so in numerical schemes. We could also maintain the ring geometry. As one example of interest, thereby one could
 investigate  collisions between the moving peaks. Universality is only expected for large $j,t$, hence in the limit of zero lattice spacing. The continuum limit of $S(j,t)$ is denoted by $S(x,t)$, $x\in\mathbb{R}$. With these steps we arrive at the mode-coupling equation
 \begin{eqnarray}\label{B.20}
&&\hspace{-53pt}  \partial_t S_{\alpha\beta}(x,t)= \sum^n_{\alpha'=1} \big(-c_\alpha\delta_{\alpha\alpha'}\partial_x +D_{\alpha\alpha'}\partial^2_x\big) S_{\alpha'\beta}(x,t)\nonumber\\
&&\hspace{15pt} + \int^t_0 ds \int_{\mathbb{R}} dy   \partial^2_y M_{\alpha\alpha'}(y,s) S_{\alpha'\beta}(x-y,t-s)
\end{eqnarray}
with the memory kernel
\begin{equation}\label{B.21}
M_{\alpha\alpha'}(x,t)= 2\sum^n_{\beta',\beta'',\gamma',\gamma''=1} G^\alpha_{\beta'\gamma'} G^{\alpha'}_{\beta''\gamma''} S_{\beta'\beta''}(x,t) S_{\gamma'\gamma''}(x,t)\,.
\end{equation}


\subsection{Fourier transform conventions}\label{B.c}

Our equations of motion are written mostly in terms of the index $j$ and its continuum approximation $x$. For numerical simulations, and also for the asymptotic analysis, it is convenient to use Fourier space.  
We list here the conventions used throughout.

We naively discretize~(\ref{B.20}) in replacing $x\in\mathbb{R}$ by $j\in\mathbb{Z}$, $\partial_x f(x)$ by $\frac{1}{2}( f(j+1)-f(j-1))$, and $\partial^2_x f(x)$ by $f(j+1)-2f(j)+f(j-1)$. Then~(\ref{B.20}) becomes
 \begin{eqnarray}\label{B.24}
&&\hspace{-20pt}  \partial_t S_{\alpha\beta} (j,t)= -c_\alpha \tfrac{1}{2}\big(S_{\alpha\beta}(j+1,t)-S_{\alpha\beta}(j-1,t)\big) 
\\
&&\hspace{-10pt} +\sum^n_{\alpha'=1}\Big(D_{\alpha\alpha'} \big(S_{\alpha'\beta}(j+1,t) -2S_{\alpha'\beta}(j,t)+S_{\alpha'\beta}(j-1,t)\big)  \nonumber\\
&&\hspace{-10pt}+\int^t_0 ds \sum_{j'\in\mathbb{Z}} \big(M_{\alpha\alpha'}(j' +1,s)  -2 M_{\alpha\alpha'}(j',s)+ M_{\alpha\alpha'}(j'-1,s)\big)  S_{\alpha'\beta}(j-j',t-s)\Big)\,.\nonumber
\end{eqnarray}
For finite $N$, $j=1,\ldots,N$, periodic boundary conditions are understood.

We adopt the standard discrete Fourier transform,
\begin{equation}\label{B.25}
\hat{f}(k)= \sum_{j\in\mathbb{Z}} f(j) \mathrm{e}^{-\mathrm{i} 2\pi jk}\,,\quad f(j)=\int^{1/2}_{-1/2} dk \hat{f}(k) \mathrm{e}^{\mathrm{i}2\pi jk}\,.
\end{equation}
$\hat{f}$ is one-periodic and the standard Brillouin zone is $k\in[-\frac{1}{2},\frac{1}{2}]$. For finite $N$,
\begin{equation}\label{B.26}
\hat{f}(k)= \sum^N_{j=1} f(j) \mathrm{e}^{-\mathrm{i}2\pi jk}\,,\quad f(j)=\frac{1}{N}\sum_{k\in[0,1]_N} \hat{f}(k) \mathrm{e}^{\mathrm{i}2\pi jk} 
\end{equation}
with $[0,1]_N = \{k|k=0,1,\ldots,N-1\}$.
In Fourier space~(\ref{B.24}) becomes
\begin{eqnarray}\label{B.27}
&&\hspace{-53pt}  \partial_t \hat{S}_{\alpha\beta}(k,t)= -\mathrm{i} c_\alpha \sin (2\pi k) \hat{S}_{\alpha\beta}(k,t) -2(1-\cos(2\pi k)) \nonumber\\
&&\hspace{15pt} \sum^n_{\alpha'=1} \Big(D_{\alpha\alpha'}\hat{S}_{\alpha'\beta}(k,t)+\int^t_0 ds  \hat{M}_{\alpha\alpha'}(k,s) \hat{S}_{\alpha'\beta}(k,t-s)\Big)\,,
\end{eqnarray}
where
\begin{equation}\label{B.28}
\hat{M}_{\alpha\alpha'} (k,s)= 2 \sum^n_{\beta,\beta',\gamma,\gamma'=1} G^\alpha_{\beta\gamma} G^{\alpha'}_{\beta'\gamma'} \int^{1/2}_{-1/2} dq \hat{S}_{\beta\beta'}
(k-q,s) \hat{S}_{\gamma\gamma'} (q,s)\,.
\end{equation}
Since $S$ refers here to the normal mode covariance, the initial conditions are 
\begin{equation}\label{B.29}
S_{\alpha\beta} (j,0)=\delta_{\alpha\beta} \delta_{j0}\,,\quad \hat{S}_{\alpha\beta}
 (k,0)=\delta_{\alpha\beta}\,.
\end{equation}

Correspondingly we use for the continuum Fourier transform
\begin{equation}\label{B.30}
\hat{f}(k)= \int dx f(x) \mathrm{e}^{-\mathrm{i}2\pi xk}\,,\quad f(x)=\int dk \hat{f}(k) \mathrm{e}^{\mathrm{i}2\pi xk}\,.
\end{equation}
Convolutions come with no extra factor of $\pi$, while $\partial_x f(x)$ is transformed to  $\mathrm{i}2\pi k \hat{f}(k)$. Also $\hat{f}(0)=\int dx f(x)$.


\section{Appendix: Levy asymptotics for the heat mode}\label{app.C}
\setcounter{equation}{0}

The starting equation reads
\begin{equation}\label{C.1}
\partial_t \hat{f}(k,t)=- (2\pi k\lambda_0)^2 \int^t_0 ds \hat{f}(k,t-s) \int_\mathbb{R} dq \hat{g}(k-q,s)\hat{g}(q,s)\,,
\end{equation}
$\hat{f}(k,0)=1$. The function $\hat{g}$ is assumed to have the scaling form
\begin{equation}\label{C.2}
\hat{g}(k,t)= \mathrm{e}^{\mathrm{i}2\pi kct} \hat{g}_0\big((\lambda_\mathrm{s} t)^\beta k\big)\,,\quad \hat{g}_0(0)=1\,,
\end{equation}
with $0<\beta<1$, $\lambda_\mathrm{s} >0$. We claim that with $\gamma=1+\beta$
\begin{equation}\label{C.3}
\lim_{k\to 0} \hat{f}(k,|k|^{-\gamma}t)= \mathrm{e}^{-\lambda_{\beta} t}\,,
\end{equation}
where the non-universal coefficient, $\lambda_\beta$, reads
\begin{equation}\label{C.4}
\lambda_\beta= (2\pi \lambda_0)^2(\lambda_\mathrm{s})^{-\beta} \int^\infty_0 dt t^{-\beta} 
\mathrm{e}^{\mathrm{i}2\pi \mathrm{sgn}(k)ct} \int_\mathbb{R}   dk |\hat{g}_0(k)|^2\,.
\end{equation}

To prove~(\ref{C.3}),~(\ref{C.4}) we look for a scaling function $\hat{h}$ such that, for $t$ of order $|k|^\gamma$,
\begin{equation}\label{C.5}
\hat{f}(k,t )= \hat{h}(|k|^\gamma t)\,.
\end{equation}
We first set $k> 0$. Inserting in~(\ref{C.1}) one obtains
\begin{equation}\label{C.6}
 \hat{h}'(|k|^\gamma t) |k|^\gamma = - (2\pi k \lambda_0)^2    \int^t_0 ds \hat{h}(|k|^\gamma (t -s)) 
 (\lambda_\mathrm{s}s)^{-\beta} \mathrm{e}^{\mathrm{i}2\pi kcs} \int_\mathbb{R} dq |\hat{g}_0(q)|^2\, .            
 \end{equation}
We substitute $w = |k|^\gamma t$ and $s$ by $s/|k|$. Then
\begin{eqnarray}\label{C.7}
&&\hspace{-20pt}\hat{h} '(w) |k|^\gamma \\
&&\hspace{0pt}= - (2\pi \lambda_0)^2|k|^{(1+\beta)}   \int^{|k|^{(-\gamma +1)}w}_0 ds \hat{h}(w - |k|^{\gamma - 1}s) 
 s^{-\beta} \mathrm{e}^{\mathrm{i}2\pi cs}  (\lambda_\mathrm{s})^{-\beta}\int_\mathbb{R} dq |\hat{g}_0(q)|^2\, .  \nonumber          
 \end{eqnarray}
Taking the limit $k \to 0$,  $\hat{h}$ can be taken out of the $s$-integral with the result $\hat{h}' = -\lambda\hat{h}$, which
yields~(\ref{C.3}). In repeating the same computation for $k< 0$, only the sign in the oscillating factor changes. 

Working out the integration in (\ref{C.4}), the scaling function at $t=1$ becomes
\begin{equation}\label{C.8}
\hat{h}(|k|^\gamma)= \exp\big[-\lambda_\beta |k|^\gamma \big(1 - \mathrm{i}\,\mathrm{sgn}(k)\tan(\tfrac{1}{2}\pi\gamma) \big)\big]\,, 
\end{equation}
where
\begin{equation}\label{C.9}
\lambda_\beta = (2 \pi \lambda_0)^2(\lambda_\mathrm{s})^{-\beta} (2\pi c)^{-1+\beta} 
\tfrac{1}{2}\pi \frac{1}{\Gamma(\beta)} \frac{1}{\cos(\tfrac{1}{2}\pi \beta)}
 \int_\mathbb{R}   dk |\hat{g}_0(k)|^2\,.
\end{equation}
One recognizes (\ref{C.9}) as the Fourier transform of asymmetric $\alpha$-stable law with $\alpha = \gamma$.
The asymmetry parameter is at its limiting allowed value. This implies a decay in position space as $|x|^{-\gamma -1}$ to the left and
as  $\exp[-|x|^{\gamma/(\gamma-1)}]$ to the right, see \cite{UZ99} for the asymptotics. In the context of fluctuating hydrodynamics this feature implies that outside the sound cone the correlations are suppressed faster than exponential.


\section{Appendix: Anharmonic chains and one-dimensio\-nal fluids}\label{app.D}
\setcounter{equation}{0}

One-dimensional fluids are governed by the hamiltonian
\begin{equation}\label{D.1}
H_\mathrm{f}=\sum^N_{j=1} \tfrac{1}{2}p^2_j +\tfrac{1}{2}\sum^N_{i\neq j=1} V(q_i-q_j)
\end{equation}
with an even interaction potential, $V(x)=V(-x)$. $V$ must be chosen such that the system is thermodynamically stable, to say $V$ may have a hard core and $V$ decays sufficiently fast at infinity, minimal requirements being $V(x)\to 0$ as $|x|\to \infty$, faster than $- |x|^{-2}$ to avoid phase transitions, $V$ is bounded from below, and not ``too negative''. Particles move on the interval $[0,L]$ with periodic boundary conditions. Particle number, momentum, and energy are locally conserved. Their corresponding microscopic fields are written as
\begin{equation}\label{D.2}
\sum^N_{j=1}\delta(q_j-x) \,,\quad \sum^N_{j=1}\delta(q_j-x)p_j\,,\quad \sum^N_{j=1}\delta(q_j-x)
\Big(\tfrac{1}{2}p^2_j +\tfrac{1}{2}\sum^N_{i=1,i\neq j} V(q_i-q_j)\Big)\,.
\end{equation}
From the evolution equations one finds the corresponding microscopic currents as
\begin{eqnarray}\label{D.3}
&&\hspace{-15pt} J_1(x)= \sum^N_{j=1}\delta(q_j-x)p_j\,,\nonumber\\
&&\hspace{-15pt} J_2(x)= \sum^N_{j=1}\delta(q_j-x)p^2_j -\tfrac{1}{2}\sum^N_{i=1,i\neq j} V'(q_i-q_j)(q_i-q_j) \int^1_0 dx \delta (\lambda q_i +(1-\lambda) q_j-x)\,,\nonumber\\
&&\hspace{-15pt} J_3(x)= \sum^N_{j=1}\delta(q_j-x)p_j\Big(\tfrac{1}{2}p^2_j +\tfrac{1}{2}\sum^N_{i=1,i\neq j} V(q_i-q_j)\Big)\,,\nonumber\\
&&\hspace{30pt} +\tfrac{1}{2}\sum^N_{i,j=1}\tfrac{1}{2}(p_i+p_j) V'(q_i-q_j)(q_i-q_j) \int^1_0 d\lambda \delta (\lambda q_i +(1-\lambda) q_j-x)\,.
\end{eqnarray}
Even without averaging, the densities from~(\ref{D.2}) and the currents from~(\ref{D.3}) satisfy a system of conservation laws.

If the fluid starts and approximately remains in local equilibrium, one can equilibrium average the microscopic conservation laws and arrives at the Euler equations of a one-dimensional fluid. The hydrodynamic fields are $\rho$, $\rho v$, $\rho \mathfrak{e}_\mathrm{f}$, which depend on $x,t$. $\rho$ is the local particle density, $v$ the local velocity per particle and $\mathfrak{e}_\mathrm{f}$ the local energy per particle. Then the Euler equations read
\begin{eqnarray}\label{D.5}
&&\hspace{10pt} \partial_t \rho +\partial_x (\rho v) =0\,,\nonumber\\
&&\hspace{10pt} \partial_t (\rho v)+\partial_x (\rho v^2 +p_\mathrm{f}) =0\,,\nonumber\\
&&\hspace{10pt} \partial_t (\rho \mathfrak{e}_\mathrm{f})+ \partial_x (\rho \mathfrak{e}_\mathrm{f} v +p_\mathrm{f}v)=0\,.
\end{eqnarray}
Here $p_\mathrm{f}$ is the local thermodynamic pressure, which depends on $\rho$ and the internal energy. We use the subscript ``f" to distinguish from the corresponding quantity for the anharmonic chain. To construct nonlinear fluctuating hydrodynamics we may proceed as in the main text. In particular, $G^0_{00}=0$ always and generically $G^1_{11}= -G^{-1}_{-1-1}\neq 0$. Thus up to model-dependent coefficients the overall structure remains unaltered. There is one technical difference. To determine the pressure $p_\mathrm{f}$ requires a full many-body computation, in general. One has then to resort to series expansions and Monte Carlo techniques. In particular, to obtain with sufficient accuracy the coupling constants $G$ will be more costly than for the chain.

If one introduces a hard core,  the ordering of particles is preserved and by a suitable choice of parameters one can achieve that only the nearest neighbor contribution to the potential term remains. An explicit example is
\begin{equation}\label{D.6}
V_{\mathrm{hc}}(x)=
\begin{cases}
  \infty\,, & |x|\leq a/2\,, \\
  \mathrm{``arbitrary}"\,, & a/2\leq |x|\leq a\,, \\
  0\,, & a\leq |x|\,. \\
\end{cases}
\end{equation}
Then, on the restricted configuration space,
\begin{equation}\label{D.7}
\tfrac{1}{2}\sum^N_{i=1,i\neq j} V_{\mathrm{hc}}(q_i-q_j) = \sum^N_{j=1} V_{\mathrm{hc}}(q_{j+1}-q_j)
\end{equation}
 Thus for $V_{\mathrm{hc}}$ one may adopt the field theory point of view with coupling only between nearest neighbors.

At second thoughts one is puzzled, since the hydrodynamic equations for a fluid look different from the ones of the field theory. One reason is that
$\rho=\ell^{-1}$. More importantly $\partial_x$ above refers to per unit length while the $\partial_x$ in~(\ref{15}),~(\ref{16}) refers to the particle label. Still, because it is the same physical system, there must be a transformation which converts~(\ref{D.5}) to~(\ref{15}),~(\ref{16}). I did not find a discussion in the literature and hence explain how the transformation is done.

First one has to properly distinguish between ``$x$'' in~(\ref{15}) and in~(\ref{D.6}). To do so, only in this section, we replace $x$ in~(\ref{15}) by $y$. Let us define
\begin{equation}\label{D.8}
\int^x_{-\infty} dx' \rho(x',t)= y_t(x)\,,\quad \int^y_{-\infty} dy' \ell(y',t)= x_t(y)\,.
\end{equation}
$y_t$ is the inverse function of $x_t$. It follows that
\begin{equation}\label{D.9}
  \rho(x,t) \ell(y,t)=1\,,
\end{equation}
where we used
\begin{equation}\label{D.10}
\partial_x y_t(x)=\rho(x,t)\,,\quad \partial_y x_t(y)=\ell(y,t)\,.
\end{equation}
If in a single equation, as~(\ref{D.9}), both $x$ and $y$ appear, they are understood to be related as in~(\ref{D.8}). Using the conservation of mass one obtains
\begin{equation}\label{D.11}
\partial_t y_t(x)=-\rho(x,t) v(x,t)\,.
\end{equation}
Other quantities flow along, \textit{i.e.}
\begin{equation}\label{D.11a}
v(x,t)= u(y,t)\,,\quad \mathfrak{e}_\mathrm{f}(x,t)=\mathfrak{e}(y,t)\,,\quad p_\mathrm{f}(x,t)=p(y,t)\,.
\end{equation}

We first discuss mass conservation and write
\begin{eqnarray}\label{D.12}
&&\hspace{-19pt} 0=\partial_t \rho + \partial_x(\rho v)=\partial_t\big(\ell(y_t(x),t)^{-1}\big)+\rho \partial_x u(y_t(x),t)+v \partial_x\big(\ell(y_t(x),t)^{-1}\big)\nonumber\\
&&\hspace{-10pt} =-\ell^{-2} (\partial_y\ell)(\partial_t y_t(x))-\ell^{-2} \partial_t \ell +\rho\big(\partial_y u(y,t)\big)\big(\partial_x y_t(x)\big) -v\ell^{-2}\big(\partial_y \ell(y,t)\big)\big(\partial_x y_t(x)\big)\nonumber\\
&&\hspace{-10pt} =\ell^{-3} v \partial_y\ell-\ell^{-2} \partial_t\ell+\rho^2 \partial_y u -v\ell^{-3}\partial_y \ell\nonumber\\
&&\hspace{-10pt} = -\ell^{-2} (\partial_t \ell - \partial_y u)\,,
\end{eqnarray}
as to be shown. We turn to momentum conservation
\begin{eqnarray}\label{D.13}
&&\hspace{-19pt} 0=\partial_t (\rho v) + \partial_x(\rho v^2) + \partial_x p_\mathrm{f}= \rho \partial_t v + \rho v \partial_x v +\partial_x p_\mathrm{f} \nonumber\\
&&\hspace{-10pt} =\rho \partial_t u (y_t(x),t) + \rho v\partial_x u(y_t(x),t) +\partial_x p_\mathrm{f}\nonumber\\
&&\hspace{-10pt} = -\rho (\partial_y u)\rho v +\rho \partial_t u+\rho v (\partial_y u) \rho +\partial_x p_\mathrm{f}\nonumber\\
&&\hspace{-10pt} = \rho \partial_t u + \partial_x p_\mathrm{f}\,.
\end{eqnarray}
Now for any pair of functions $g$, $g_\mathrm{f}$ such that $g_\mathrm{f}(x,t)=g(y,t)$, one has
\begin{equation}\label{D.14}
\partial_x g_\mathrm{f}(x,t)= \partial_x g\big(y_t(x),t\big)=\big(\partial_y g(y,t)\big)\big(\partial_x y_t(x)\big)=\big(\partial_y g(y,t)\big) \rho(x,t)\,.
\end{equation}
From~(\ref{D.13}) one concludes that
\begin{equation}\label{D.15}
\partial_t u +\partial_y p=0\,.
\end{equation}
Finally we turn to energy conservation, with similar steps as in~(\ref{D.15})
\begin{eqnarray}\label{D.16}
&&\hspace{-19pt} 0=\partial_t (\rho \mathfrak{e}_\mathrm{f}) + \partial_x(\rho \mathfrak{e}_\mathrm{f} v) + \partial_x (vp_\mathrm{f}) \nonumber\\
&&\hspace{-10pt} =\rho \partial_t \mathfrak{e}_\mathrm{f} + \rho v\partial_x \mathfrak{e}_\mathrm{f} +\partial_x (v p_\mathrm{f})\nonumber\\
&&\hspace{-10pt} = \rho \partial_t \mathfrak{e} +\partial_x (v p_\mathrm{f})=  \rho \big(\partial_t \mathfrak{e}+\partial_y (p u)\big)\,,
\end{eqnarray}
where we used~(\ref{D.14}) with $g_\mathrm{f}=v p_\mathrm{f}$, $g= up$. The pressure $p_\mathrm{f}$ depends on $x,t$ through $\rho$ and $\mathsf{e}_\mathrm{f}$ and the pressure $p$ depends on $y,t$ through $\ell$ and $\mathsf{e}$.

\end{appendix}

\end{document}